\documentclass[thmsa,12pt]{report}
\usepackage{amssymb}
\usepackage[brazil]{babel}

\begin{document}

\label{capa}
\thispagestyle{empty}

\vspace{2.5cm}

\centerline{\large \sc Tese de Doutorado}

\vspace{3cm}

\centerline{\Large \bf Estudo da Consist\^encia do M\'etodo dos Projetores Simpl\'eticos }
\centerline{\Large \bf atrav\'es de Modelos de Gauge Multidimensionais}

\vspace{4cm}

\centerline{\large Marco Antonio dos Santos}

\vspace{6cm}

\centerline{Centro Brasileiro de Pesquisas F\'{\i}sicas}

\bigskip

\centerline{Rio de Janeiro}

\bigskip

\centerline{Dezembro de 2001}%

\newpage
\thispagestyle{empty}

~\vspace{18cm}

\hspace{5cm} Ao meu triunvirato , Alda, Simone e Francine.

\bigskip\bigskip

\hspace{5cm} Ao {\bf Prof. C. M. do Amaral.}

\newpage

\thispagestyle{empty}
\centerline{\bf AGRADECIMENTOS}

\bigskip

Ao  Jos\'{e}  Helay\"{e}l, que impressiona aos que o conhecem por sua dimens%
\~{o}es, humanas e profissionais, e que tive a felicidade em ter como
orientador e amigo.

Ao  Marquinho, n\~{a}o apenas meu co-orientador, mas um companheiro no
melhor sentido da palavra, com quem criei uma das amizades mais valiosas em
toda a minha vida.

Ao Ion, por todo o aux\'{i}lio inestim\'{a}vel que prestou neste trabalho, e
por tamb\'{e}m enriquecer tanto minha curta e preciosa lista de grandes
amigos.

Ao Prof. Caride, pelo cr\'{e}dito com que nos acolheu na CCP/CBPF,
possibilitando a realiza\c{c}\~{a}o deste trabalho.

A Miriam pelo carisma e pela impressionante efici\^encia com que secretaria a CFC.

Ao maravilhoso Grupo de F\'{i}sica Te\'{o}rica da UCP, onde tive o ambiente
e a hospitalidade fundamentais para a realiza\c{c}\~{a}o deste trabalho, com
destaque especial aos novos parceiros \'{A}lvaro e  Humberto, e ao velho
camarada  Colatto.

Aos velhos e bons companheiros do CBPF, em especial ao meu \ ''irm\~{a}o ''
Paulo Israel, por raz\~{o}es t\~{a}o variadas quanto essenciais.

Ao grande colega V\'{a}lter, da UFRRJ, parceiro constante nas dificuldades
que enfrentamos hoje na Universidade  para realizar nosso trabalho.

Ao Departamento de F\'{i}sica da UFRRJ, que permitiu a realiza\c{c}\~{a}o
deste trabalho.

Aos irm\~{a}os Beto e Miro Maiworm, que t\~{a}o bem e gentilmente t\^{e}m
recebido os companheiros em Petr\'{o}polis, colaborando para o ambiente saud%
\'{a}vel em que fazemos nosso trabalho.

Ao Prof. Carlos M\'{a}rcio do Amaral (in memoriam), mentor original
desta tese.

\newpage

\thispagestyle{empty}
\begin{abstract}

Neste trabalho, adotamos o tratamento can\^onico para estudar a quantiza\c c\~ao de
mo\-de\-los de gauge de categorias bastante distintas, e em diversas dimens\~oes. 
O m\'{e}todo dos projetores simpl\'{e}ticos \'{e} utilizado e sua consist\^{e}ncia 
verificada para identificar o espa\c{c}o de fase reduzido em cada caso.

\end{abstract}

\newpage 
~\vspace{6cm}
\thispagestyle{empty} 
\centerline{\bf Abstract}

\bigskip

In this work, we analyse several classes of gauge models from the canonical
viewpoint. The symplectic projector method is applied, and its consistency in the
identification of the reduced phase space is checked for each case.

\newpage

\tableofcontents
\thispagestyle{empty}

\chapter{\protect\bigskip {\sc Introdu\c{c}\~{a}o}}

Uma caracter\'{\i}stica comum a todas as formula\c{c}\~{o}es te\'{o}ricas
que descrevem as intera\c{c}\~{o}es fundamentais da Natureza \'{e} a
exist\^{e}ncia de uma simetria interna chamada simetria de gauge:

\begin{center}
Inter. Fundamental $~~\Longrightarrow~~$ Simetria de Gauge.
\end{center}

\noindent Esta simetria introduz aspectos que dificultam o tratamento destas
teorias, uma vez que da\'{\i} decorre uma singularidade na Lagrangeana,
refletida no aparecimento de um Hessiano nulo:

\begin{center}
Simetria de Gauge $~~\Longrightarrow~~$ $\mathcal{L}$ singular
\end{center}

\noindent Por sua vez, uma teoria com Lagrangeana singular implica, ao se
tentar construir um formalismo Hamiltoniano, na exist\^{e}ncia de
v\'{\i}nculos entre as coordenadas do espa\c{c}o de fase:

\begin{center}
$\mathcal{L}$ singular $~~\Longrightarrow~~$ $\phi _{m}\left( q,p\right) =0$
\end{center}

\noindent Este quadro tem despertado um grande interesse nos \'{u}ltimos 50
anos, visto que s\~{a}o evidentes as dificuldades em se estabelecer um
formalismo de quantiza\c{c}\~{a}o can\^{o}nica neste caso.

Realmente, um progresso not\'{a}vel foi alcan\c{c}ado nas \'{u}ltimas duas
d\'{e}cadas com o chamado formalismo BRST \cite{teitelboim}. Verificou-se,
em linhas gerais, que o tratamento deste problema poderia ser feito de
maneira eficaz ao se estender o espa\c{c}o de fase, via introdu\c{c}\~{a}o
de vari\'{a}veis extras ( com paridade de Grassmann invertida), e obter uma
simetria mais geral, que permite a recupera\c{c}\~{a}o da estrutura
can\^{o}nica desejada. As quantidades cl\'{a}ssicas e qu\^{a}nticas s\~{a}o
vistas como objetos no grupo de cohomologia de um operador nilpotente que
codifica a simetria do sistema. A superf\'{\i}cie ``f\'{\i}sica''
encontra-se embebida neste espa\c{c}o ``aumentado'' e a f\'{\i}sica \'{e}
resgatada na solu\c{c}\~{a}o do problema cohomol\'{o}gico. Apesar deste
m\'{e}todo, em suas v\'{a}rias formula\c{c}\~{o}es, tratar com todas as situa%
\c{c}\~{o}es de uma maneira marcadamente eficiente, permitindo a solu\c{c}%
\~{a}o de problemas complexos como a identifica\c{c}\~{a}o de anomalias,
termos de Schwinger e renormaliza\c{c}\~{a}o, perde-se a\'{\i} um tanto do
quadro intuitivo do espa\c{c}o f\'{\i}sico.

Por outro lado, a solu\c{c}\~{a}o no sentido oposto, qual seja, o de \textit{%
reduzir }o espa\c{c}o original, eliminando-se as vari\'{a}veis esp\'{u}rias
cuja exist\^{e}ncia os v\'{\i}nculos acusam, e obtendo assim o espa\c{c}o
f\'{\i}sico correspondente, n\~{a}o tem obtido o mesmo progresso. A
dificuldade fundamental neste sentido seria a de n\~{a}o ser trivial a
elimina\c{c}\~{a}o de vari\'{a}veis atrav\'{e}s da express\~{a}o dos
v\'{\i}nculos no caso geral. \'{E} precisamente nesta dire\c{c}\~{a}o que o
presente trabalho de tese caminha.

O m\'{e}todo que apresentaremos, e que doravante chamaremos de ``m\'{e}todo
dos projetores simpl\'{e}ticos (MPS), nasceu de um trabalho original do
Prof. C. M. do Amaral \cite{cma}, tratando sistemas cl\'{a}ssicos discretos,
n\~{a}o- relativ\'{\i}sticos, sujeitos a um conjunto de v\'{\i}nculos
hol\^{o}nomos independentes no espa\c{c}o de configura\c{c}\~{o}es. Ao
identificar a hipersuperf\'{\i}cie onde se realiza a din\^{a}mica com aquela
definida pelos v\'{\i}nculos, Amaral construiu um projetor que define,
localmente, um conjunto de coordenadas linearmente independentes no espa\c{c}%
o tangente isomorfo \`{a} superf\'{\i}cie dos v\'{\i}nculos. Este conjunto
de coordenadas, livre de v\'{\i}nculos, serve como base para a descri\c{c}%
\~{a}o da din\^{a}mica sobre a superf\'{\i}cie.

Foi seguindo sugest\~{a}o sua que nos pusemos a trabalhar de forma
an\'{a}loga na constru\c{c}\~{a}o de uma estrutura geometricamente
equivalente no espa\c{c}o de fase de teorias de gauge. N\~{a}o foi
dif\'{\i}cil perceber que sua estrat\'{e}gia geom\'{e}trica poderia ser
utilizada no espa\c{c}o de fase (\cite{cma e pitanga},\cite{pitanga}) com a
\'{u}nica e severa restri\c{c}\~{a}o de que os v\'{\i}nculos presentes se
enquadrem na categoria de v\'{\i}nculos de segunda-classe, na terminologia
de Dirac \cite{dirac}. O que em teorias de gauge significa exatamente a condi%
\c{c}ao de realizar o \textit{gauge-fixing }desde o in\'{\i}cio (\cite
{sudarsham} ,\cite{sundermeyer} ,\cite{teitelboim}). Desta maneira, pudemos
testar o MPS em diversos exemplos de teorias de gauge, desde situa\c{c}%
\~{o}es cl\'{a}ssicas como o eletromagnetismo em 4-$D$ \cite{qed} at\'{e}
problemas mais atuais como cordas n\~{a}o-comutativas \cite{mmi}. Apesar do
fato de ser menos poderoso que o m\'{e}todo BRST e mais restrito em suas
aplica\c{c}\~{o}es, o MPS \'{e} bem mais simples de tratar nas v\'{a}rias
situa\c{c}\~{o}es em que n\~{a}o se esteja interessado em manter toda a
generalidade de BRST. Em alguns casos os resultados concordaram com aquilo
que j\'{a} se conhecia, enquanto que em outros o MPS ajudou na investiga\c{c}%
\~{a}o do conte\'{u}do f\'{\i}sico dos modelos em quest\~{a}o.

Este trabalho estar\'{a} organizado da seguinte maneira: no Cap\'{\i}tulo 2
faremos uma breve revis\~{a}o dos conceitos b\'{a}sicos utilizados por
Amaral \cite{cma}, a extens\~{a}o ao espa\c{c}o de fase e \`{a}s teorias de
gauge, assim como algumas considera\c{c}\~{o}es gerais a respeito do
projetor e a estrutura do espa\c{c}o f\'{\i}sico. No Cap\'{\i}tulo 3
trataremos de sistemas simples que podem bem ilustrar o MPS (\cite{c-lee} , 
\cite{revmmi}). Em seguida, no Cap\'{\i}tulo 4, analisaremos alguns problemas
de teorias de gauge onde pudemos aplicar o MPS e obter resultados
interessantes (\cite{qed}, \cite{helayel}, \cite{alvaro}). Finalmente, o
Cap\'{\i}tulo 5 det\'{e}m-se na aplica\c{c}\~{a}o do m\'{e}todo a teorias de
campos para objetos estendidos, tratando especificamente de uma corda
bos\^{o}nica com extremidades ligadas a D-branas \cite{mmi}. Conclu\'{\i}mos
com as Considera\c{c}\~{o}es Finais e Futuras Perspectivas, discutidas no
Cap\'{\i}tulo 6.

\chapter{{\sc Fundamentos Te\'oricos}}

Neste cap\'{\i}tulo mostraremos como, partindo dos argumentos originais do
Prof. C. M. do Amaral \cite{cma}, constru\'{\i}mos o projetor simpl\'{e}tico
para teorias de campos no espa\c{c}o de fase. A constru\c{c}\~{a}o,
puramente geom\'{e}trica, toma por base a natureza simpl\'{e}tica da
geometria presente. Conex\~{o}es com o tratamento usual, de Dirac \cite
{dirac}, s\~{a}o tamb\'{e}m discutidas, assim como algumas propriedades
gerais recentemente estudadas \cite{revmmi}.

\section{{\sc Espa\c cos de Configura\c c\~oes}}

Consideremos como ponto de partida um sistema discreto para o qual o espa\c{c%
}o de configura\c{c}\~{o}es $E^{n}$ \'{e} Euclidiano e isomorfo a $R^{n}$.
As coordenadas Cartesianas globais s\~{a}o $x^{\nu }$, onde $\nu =1,\ldots
,n $. Por hip\'{o}tese o sistema se encontra sujeito a um conjunto
m\'{\i}nimo de $r$ v\'{\i}nculos hol\^{o}nomos independentes. A
superf\'{\i}cie de v\'{\i}nculos \'{e} descrita pelo seguinte conjunto de
equa\c{c}\~{o}es independentes 
\begin{equation}
\Sigma ~:~~~\phi ^{I}(x^{\nu })=0,  \label{1}
\end{equation}
onde $I=1,2,\ldots ,r$. A superf\'{\i}cie $\Sigma $ \'{e} geom\'{e}trica e
independente da din\^{a}mica. Portanto, os objetos geom\'{e}tricos
relacionados a ela s\~{a}o invariantes frente a evolu\c{c}\~{a}o do sistema.
Entretanto, a din\^{a}mica se realiza sobre $\Sigma $ e queremos encontrar
um sistema de coordenadas local sobre ela e livre de v\'{\i}nculos.

O primeiro ponto a ser notado \'{e} que \'{e} poss\'{\i}vel construir um
sistema de coordenadas local em cada ponto $x\in E^{n}$ de coordenadas
globais $x^{\nu }$ se um campo matricial regular $M(x^{\nu })$ \'{e} dado.
Como a constru\c{c}\~{a}o \'{e} puramente geom\'{e}trica, os elementos de $%
M(x^{\nu })$ devem ser independentes do tempo. Se $\mathbf{e}_{\mu }$ \'{e} a 
base ortogonal de $E^{n}$ associada \`{a}s
coordenadas Cartesianas globais, a base local $\mathbf{f}_{a}(x)$, onde $%
a=1,2,\ldots ,n$ , associada ao sistema de coordenadas em $P$ \'{e} dada por 
\begin{equation}
\mathbf{f}_{a}(x)=M_{a}^{\mu }(x)\mathbf{e}_{\mu }.  \label{2}
\end{equation}
Em geral, $M_{a}^{\mu }(x)$ definir\'{a} coordenadas locais tamb\'{e}m em $%
\Sigma $. A configura\c{c}\~{a}o do sistema no instante $t$ ser\'{a} dada
por um vetor posi\c{c}\~{a}o $\mathbf{x}(t)$ , no espa\c{c}o de configura\c{c%
}\~{o}es, de um ponto da trajet\'{o}ria contida em $\Sigma $.

A fim de projetar na superf\'{\i}cie de v\'{\i}nculos, escolhemos uma base
local espec\'{\i}fica em cada ponto $x\in \Sigma $ separando o espa\c{c}o
tangente $T_{x}^{n}(E)$ \ a $\ E^{n}$ em $x$ na soma direta $%
T_{x}^{n-r}(\Sigma )\oplus N_{x}^{r}(\Sigma )$. Aqui,$T_{x}^{n-r}(\Sigma )$
\'{e} o espa\c{c}o tangente a $\Sigma $ em $x$ e $N_{x}^{r}(\Sigma )$ \'{e}
o espa\c{c}o normal. A base local \'{e} ent\~{a}o separada como $\mathbf{%
f_{a}}(x)=\{\mathbf{n}_{I}(x),\mathbf{t}_{j}(x)\}$, onde $\mathbf{n}_{I}(x)$
forma a base normalizada de $N_{x}^{r}(\Sigma )$ enquanto $\mathbf{t}_{j}$, $%
j=r+1,\ldots ,n$ , \'{e} a base no espa\c{c}o tangente $T_{x}^{n-r}(\Sigma )$%
.

A base $\mathbf{n}^{I}(x)$ pode ser constru\'{\i}da em termos dos
v\'{\i}nculos $\phi ^{I}(x)$ ; sejam 
\begin{equation}
C_{\mu }^{I}\equiv \frac{\partial \phi ^{I}(x)}{\partial x^{\mu }}  \label{3}
\end{equation}
as componentes do vetor $\mathbf{C}^{I}$, normal \`{a} superf\'{\i}cie de
v\'{\i}nculos. Como os v\'{\i}nculos s\~{a}o, por hip\'{o}tese, linearmente
independentes, o conjunto de vetores unit\'{a}rios 
\begin{equation}
\mathbf{n}^{I}(x)=\frac{\mathbf{C}^{I}\left( x\right) }{\left| \mathbf{C}%
^{I}\left( x\right) \right| }  \label{4}
\end{equation}
forma uma base n\~{a}o - ortogonal no espa\c{c}o $N_{x}^{r}(\Sigma ).$ Estes
vetores geram a m\'{e}trica $g^{IJ}(x)$, que pode ser expressa em termos da
m\'{e}trica global $g^{\mu \nu }$ como 
\begin{equation}
g^{IJ}=g^{\mu \nu }\partial _{\mu }\phi ^{I}\partial _{\nu }\phi ^{J}\;.
\label{5}
\end{equation}

De posse destes objetos, \'{e} f\'{a}cil construir um projetor que projete
os vetores de $E^{n}$ na vizinhan\c{c}a de $x$ \cite{cma}. Uma maneira de
ver isso \'{e}, usando a nota\c{c}\~{a}o bra-ket de Dirac, onde $%
\left\langle \mathbf{n}_{K}\right| $ est\'{a} na base can\^{o}nica do espa\c{%
c}o cotangente $T_{x}^{\ast (n-r)}$ associado a $\left| \mathbf{n}%
_{J}\right\rangle $, escrever o projetor $\Lambda (x)$ como 
\begin{equation}
\Lambda (x)=\mathbf{1}-g^{JK}\left| \mathbf{n}_{J}\right\rangle \left\langle 
\mathbf{n}_{K}\right| .  \label{7}
\end{equation}
Realmente, escrevendo o vetor gen\'{e}rico $\mathbf{V}$ na base local como $%
\mathbf{V=}V^{I}\left| \mathbf{n}_{I}\right\rangle +V^{j}\left| \mathbf{t}%
_{j}\right\rangle$ vemos que 
\begin{eqnarray*}
\Lambda V &=&\left[ \mathbf{1}-g^{JK}\left| \mathbf{n}_{J}\right\rangle
\left\langle \mathbf{n}_{K}\right| \right] \left( V^{I}\left| \mathbf{n}%
_{I}\right\rangle +V^{j}\left| \mathbf{t}_{j}\right\rangle \right) \\
&=& \mathbf{V}^{I}\left| \mathbf{n}_{I}\right\rangle +V^{j}\left| \mathbf{t}%
_{j}\right\rangle -g^{JK}\left| \mathbf{n}_{J}\right\rangle \left\langle 
\mathbf{n}_{K}\right| \left( V^{I}\left| \mathbf{n}_{I}\right\rangle
+V^{j}\left| \mathbf{t}_{j}\right\rangle \right) \\
&=& V^{I}\left| \mathbf{n}_{I}\right\rangle +V^{j}\left| \mathbf{t}%
_{j}\right\rangle -g^{JK}\left| \mathbf{n}_{J}\right\rangle \left\langle 
\mathbf{n}_{K}\right| V^{I}\left| \mathbf{n}_{I}\right\rangle \\
&=& V^{I}\left| \mathbf{n}_{I}\right\rangle +V^{j}\left| \mathbf{t}%
_{j}\right\rangle -V^{I}\left| \mathbf{n}_{I}\right\rangle =V^{j}\left| 
\mathbf{t}_{j}\right\rangle
\end{eqnarray*}

A partir de (\ref{7}), com (\ref{5}), podemos escrever os elementos de
matriz do projetor expl\'{\i}citamente em termos dos v\'{\i}nculos: 
\begin{equation}
\Lambda _{\nu }^{\mu }=\delta _{\nu }^{\mu }-g^{\mu \alpha }\partial
_{\alpha }\phi ^{I}g_{IJ}\partial _{\nu }\phi ^{J}  \label{8}
\end{equation}

O projetor (\ref{8}) selecionar\'{a} as componentes paralelas ao espa\c{c}o
tangente a $\Sigma $ em $x$ de qualquer vetor no espa\c{c}o de configura\c{c}%
\~{o}es. Como discutido em \cite{cma}, o espa\c{c}o tangente local \'{e}
livre de v\'{\i}nculos. A fim de encontrar a configura\c{c}\~{a}o local do
sistema em qualquer instante $t$ basta projetar o vetor de configura\c{c}%
\~{a}o $\mathbf{x}(t)$ fazendo $\Lambda $ atuar sobre ele. Os dois espa\c{c}%
os $T_{x}^{n-r}$ e $N_{x}^{r}$ s\~{a}o dados pelos autovetores do projetor
complementar $Q(x)=\mathbf{1}-\Lambda (x)$ associados aos autovalores 0 e 1,
respectivamente.

Com uma escolha apropriada da matriz $M_{a}^{\mu }(x)$, podemos separar as
componentes do vetor de configura\c{c}\~{a}o em dois conjuntos $%
\{x^{j}(t),x^{I}(t)\}$ que satisfazem as seguintes condi\c{c}\~{o}es de
contorno \cite{cma} 
\begin{equation}
\dot{x}^{I}(t)=0~~~,~~~x^{I}(t)=\phi ^{I}(x)=0,  \label{9}
\end{equation}
onde os vetores foram ordenados com $I=n-r+1,\ldots ,n$. A transforma\c{c}%
\~{a}o regular geral para coordenadas locais que satisfaz (\ref{9}) deixa
invariante a Lagrangeana e, portanto, a Lagrangeana projetada pode ser
localmente escrita como 
\begin{equation}
L^{\prime }(x^{j}(t),{\dot{x}}^{j}(t))=L(x^{\nu }(t),{\dot{x}}^{\nu }(t))
\label{10}
\end{equation}

A din\^{a}mica local \'{e} dada pelas equac\~{o}es livres de Euler-Lagrange
obtidas a partir de $L^{\prime }$ em termos de $x^{j}$ apenas, ou de $L$ com
as condi\c{c}\~{o}es de contorno (\ref{9}).

\section{{\sc O Projetor Simpl\'etico $\Lambda $ e as Vari\'aveis
F\'{\i}sicas}}

A passagem para o espa\c{c}o de fase nos obriga, em princ\'{\i}pio, a tomar
em conta a estrutura de geometria simpl\'{e}tica agora presente, e que deve
ser cuidadosamente levada em consi\-de\-ra\c{c}\~{a}o. Veremos que esta
estrutura define exatamente o tipo de v\'{\i}nculos com o qual \'{e}
poss\'{\i}vel construir um projetor em analogia com o que foi feito acima.
Antes, entretanto, vamos estabelecer alguns conceitos fundamentais na formula%
\c{c}\~{a}o can\^{o}nica de teorias de gauge.

No tratamento geral de teorias com Lagrangeano singular \cite{dirac},
revela-se a exist\^{e}ncia de rela\c{c}\~{o}es envolvendo momenta e
coordenadas ao se realizar a transforma\c{c}\~{a}o de Legendre. Estas rela\c{%
c}\~{o}es s\~{a}o chamadas de v\'{\i}nculos prim\'{a}rios da teoria. Condi\c{%
c}\~{o}es de consist\^{e}ncia, ao serem impostas sobre estes, d\~{a}o origem
a outros v\'{\i}nculos, que por isso s\~{a}o chamados de v\'{\i}nculos
secund\'{a}rios. Este processo de verifica\c{c}\~{a}o de condi\c{c}\~{o}es
de consist\^{e}ncia e gera\c{c}\~{a}o de novos v\'{\i}nculos \'{e} conhecido
na literatura como algoritmo de Dirac. Findo este processo obtem-se o
conjunto de v\'{\i}nculos ao qual o modelo se sujeita, n\~{a}o possuindo
maior relev\^{a}ncia a distin\c{c}\~{a}o entre v\'{\i}nculos prim\'{a}rios
ou secund\'{a}rios. Entretanto, uma outra distin\c{c}\~{a}o, esta sim de
car\'{a}ter fisicamente bastante relevante, tem ent\~{a}o lugar. Alguns
v\'{\i}nculos possuem par\^{e}nteses de Poisson nulos, ao menos fracamente,
na terminologia de Dirac, com todos os outros e com a Hamiltoniana. Servem
ent\~{a}o como geradores de transforma\c{c}\~{o}es can\^{o}nicas, e s\~{a}o
rotulados de v\'{\i}nculos de primeira-classe. Ser\~{a}o denotados por 
\begin{equation}
\chi _{m}\left( q,p\right) \approx 0  \;, \label{15}
\end{equation}
onde o s\'{\i}mbolo $\approx $ significa igual sobre a superf\'{\i}cie
definida pelos v\'{\i}nculos.

Todos os demais s\~{a}o chamados v\'{\i}nculos de segunda-classe, e
ser\~{a}o denotados por 
\begin{equation}
\varphi _{n}\left( q,p\right) \approx 0  \;. \label{16}
\end{equation}
Pela sua defini\c{c}\~{a}o, estes v\'{\i}nculos s\~{a}o tais que a matriz
com elementos 
\begin{equation}
\Delta _{mn}=\{\varphi _{m},\varphi _{n}\}  \label{17}
\end{equation}
\'{e} invers\'{\i}vel.

A simples presen\c{c}a de v\'{\i}nculos entre as coordenadas do espa\c{c}o
de fase induz \`{a} possibilidade de expressar algumas destas em termos de
um conjunto m\'{\i}nimo de coordenadas independentes, e este \'{e} em
ess\^{e}ncia o argumento funtamental que norteia a busca de um espa\c{c}o de
fase reduzido, ou f\'{\i}sico. Naquelas teorias em que os v\'{\i}nculos
presentes s\~{a}o todos de segunda-classe esta possibilidade, em
princ\'{\i}pio, \'{e} plenamente realiz\'{a}vel, como veremos adiante. As
dificuldades aumentam quando da presen\c{c}a de v\'{\i}nculos de
primeira-classe, como veremos a seguir.

A simetria obtida pelas transforma\c{c}\~{o}es geradas pelos v\'{\i}nculos
de primeira-classe \'{e} chamada simetria de gauge, neste contexto. As
teorias de campos de gauge s\~{a}o ent\~{a}o teorias que cont\^{e}m
v\'{\i}nculos de primeira-classe, por defini\c{c}\~{a}o. Por\'{e}m, a
exist\^{e}ncia desta simetria implica na n\~{a}o-biunivucidade na rela\c{c}%
\~{a}o entre estados f\'{\i}sicos e pontos do espa\c{c}o de fase; ou seja, a
um estado f\'{\i}sico est\~{a}o associados pontos do espa\c{c}o de fase que
diferem por uma transforma\c{c}\~{a}o can\^{o}nica gerada pelos
v\'{\i}nculos de primeira-classe. Para eliminar esta ambiguidade na descri\c{%
c}\~{a}o do sistema, um conjunto de rela\c{c}\~{o}es deve ser imposto 
\textit{ad hoc }de maneira a tornar o conjunto original de v\'{\i}nculos de
primeira-classe em um novo conjunto, estendido, agora de segunda-classe.
Este procedimento \'{e} conhecido na literatura como \textit{gauge fixing.}

Retornemos \`{a} quest\~{a}o da constru\c{c}\~{a}o do projetor no espa\c{c}o
de fase. A geometria da superf\'{\i}cie de v\'{\i}nculos \cite{teitelboim},
em se tratando de um espa\c{c}o de fase, \'{e} caracterizada pela
exist\^{e}ncia de uma dois-forma induzida, $J^{ij}$, heran\c{c}a do fato de
esta superf\'{\i}cie estar embebida em um espa\c{c}o que possui uma
dois-forma simpl\'{e}tica natural, um tensor antissim\'{e}trico
n\~{a}o-degenerado de rank-2, $\ J^{\mu \nu }$, cujas componentes, em um
sistema de coordenadas simpl\'{e}tico $\xi ^{\lambda }$ s\~{a}o fornecidas
pelos par\^{e}nteses de Poisson fundamentais: 
\begin{equation}
J^{\mu \nu }=\{\xi ^{\mu },\xi ^{\nu }\}  \label{11}
\end{equation}

Seja $2N$ a dimens\~{a}o do espa\c{c}o original, e $M$ o n\'{u}mero de
v\'{\i}nculos presentes. Como desejamos obter um espa\c{c}o de fase reduzido
que contenha apenas a f\'{\i}sica do sistema, devemos procurar condi\c{c}%
\~{o}es que nos garantam que este espa\c{c}o reduzido possua uma dois-forma
induzida invers\'{\i}vel e uma estrutura de par\^{e}nteses de Poisson bem
definida. \'{E} relativamente simples mostrar \cite{teitelboim} que esta
estrutura estar\'{a} garantida se o conjunto de v\'{\i}nculos em quest\~{a}o
\'{e} um conjunto irredut\'{\i}vel e de segunda-classe.

Embora esta seja uma restri\c{c}\~{a}o aparentemente forte, na maior parte
dos casos de interesse f\'{\i}sico ela pode ser contornada. A quest\~{a}o
envolvida aqui est\'{a} relacionada ao teorema de Darboux, que assegura que,
pelo menos localmente, \'{e} sempre poss\'{\i}vel garantir a exist\^{e}ncia
de um conjunto de condi\c{c}\~{o}es de gauge fixing que transforme um
conjunto de v\'{\i}nculos de primeira-classe em outro de segunda-classe.

Desta maneira, supondo um conjunto de v\'{\i}nculos de segunda-classe
adequadamente constru\'{\i}do, e levando em considera\c{c}\~{a}o que a
estrutura geom\'{e}trica que sustenta a cons\-tru\-\c{c}\~{a}o do projetor no
espa\c{c}o de configura\c{c}\~{o}es e no espa\c{c}o de fase \'{e}
basicamente a mesma, \'{e} f\'{a}cil ver que a extens\~{a}o natural do
projetor (\ref{8}) para o espa\c{c}o de fase, agora no cont\'{\i}nuo, \'{e}
obtida com a seguinte expres\~{a}o para o projetor simpl\'{e}tico: 
\begin{equation}
\Lambda _{\nu }^{\mu }(x,y)=\delta _{\nu }^{\mu }\,\delta ^{3}(x-y)-J^{\mu
\alpha }\int d^{\,3}\!\rho \;d^{\,3}\!\sigma \,\delta _{\alpha (x)}\varphi
^{i}(\rho )\,J_{ij}(\rho ,\sigma )\,\delta _{\nu (y)}\varphi ^{j}(\sigma )
\label{12}
\end{equation}
onde $\delta _{\alpha (x)}\varphi ^{i}(\rho )\equiv \frac{\delta \varphi
^{i}(\rho )}{\delta \xi ^{\alpha }(x)}$ , e $J_{ij}$ \'{e} a matriz inversa
daquela formada pelos Par\^{e}nteses de Poisson entre os v\'{\i}nculos, e
vem a ser a $2$-forma induzida na superf\'{\i}cie dos v\'{\i}nculos.

As vari\'{a}veis projetadas 
\begin{equation}
\xi ^{*\mu }\left( x\right) =\int dy\Lambda _{\nu }^{\mu }(x,y)\xi ^{\nu
}\left( y\right)  \label{13}
\end{equation}
s\~{a}o, em princ\'{\i}pio, independentes, livres de v\'{\i}nculos, e
possuem regras can\^{o}nicas de comuta\c{c}\~{a}o. Ser\~{a}o, daqui por
diante, chamadas de vari\'{a}veis f\'{\i}sicas.

Por analogia ao que foi discutido acima (\ref{9} , \ref{10}) \'{e} f\'{a}cil
concluir que a Hamiltoniana projetada deve ser obtida da Hamiltoniana
original via $H^{\ast }=H\left( \xi ^{\ast }\right) $; ou seja, a
Hamiltoniana f\'{\i}sica \'{e} obtida da Hamiltoniana original reescrevendo
esta em termos das vari\'{a}veis projetadas. As equa\c{c}\~{o}es de
movimento obtidas atrav\'{e}s do princ\'{\i}pio variacional neste espa\c{c}o
reduzido livre de v\'{\i}nculos s\~{a}o as equa\c{c}\~{o}es de
Hamilton-Jacobi usuais:

\begin{equation}
\dot{{\mathbf{\xi}^\ast}}=\left\{ \mathbf{\xi }^{\ast},H^{\ast}\right\}
\label{14}
\end{equation}

\section{{\sc Discuss\~oes e Conclus\~oes Gerais}}

Algumas considera\c{c}\~{o}es nos parecem dignas de notas. Em primeiro
lugar, pode-se observar a semelhan\c{c}a estrutural entre a express\~{a}o (%
\ref{12}) e a bem conhecida matriz dos Par\^{e}nteses de Dirac fundamentais:

\begin{equation}
D^{MN}=\{\xi ^{M}\;,\;\xi ^{N}\}_{D}=J^{MN}-J^{ML}J^{KN}\,\frac{\delta
\varphi _{m}}{\delta \xi ^{L}}\,\Delta _{mn}^{-1}\,\frac{\delta \varphi _{n}%
}{\delta \xi ^{K}}\;.  \label{m10}
\end{equation}
Foi proposto originalmente por Dirac utilizar estes par\^{e}nteses para
fazer a passagem aos comutadores qu\^{a}nticos. Entretanto, existe um
s\'{e}rio defeito na estrutura dos par\^{e}nteses de Dirac, qual seja, como
par\^{e}nteses estendidos, os comutadores obtidos a partir deles perdem a
estrutura de deltas de Kronecker. Uma maneira de evitar este problema \'{e}
manter par\^{e}nteses com estrutura can\^{o}nica, reduzindo a an\'{a}lise do
sistema ao subespa\c{c}o f\'{\i}sico embebido no espa\c{c}o de fase. Este
objetivo pode ser alcan\c{c}ado atrav\'{e}s da constru\c{c}\~{a}o de
operadores de proje\c{c}\~{a}o sobre a superf\'{\i}cie f\'{\i}sica, da
maneira que estamos propondo.

\'{E} f\'{a}cil ver que a seguinte rela\c{c}\~{a}o entre o projetor
geom\'{e}trico e a matriz alg\'{e}brica de Dirac existe: 
\begin{equation}
\Lambda =-DJ.  \label{m11}
\end{equation}
Esta rela\c{c}\~{a}o simples conecta dois objetos que s\~{a}o construidos
por abordagens completamente independentes.

Uma segunda observa\c{c}\~{a}o que podemos fazer refere-se ao tra\c{c}o da
matriz do projetor: considere em \ (\ref{12}) $\mu =1,..,2N$ \ e $i=1,..,M$;
sendo $J_{jk}$ definido por 
\begin{equation}
\{\varphi ^{i},\varphi ^{j}\}J_{jk}=\delta _{k}^{i}  \label{i}
\end{equation}
e como, da defini\c{c}\~{a}o dos Par\^{e}nteses de Poisson 
\begin{equation}
\{A,B\}=J^{\alpha \mu }\frac{\delta A}{\delta \xi ^{\alpha }}\frac{\delta B}{%
\delta \xi ^{\mu }},  \label{ii}
\end{equation}
podemos reescrever (\ref{i}) como 
\begin{equation}
-J^{\alpha \mu }\frac{\delta \varphi ^{i}}{\delta \xi ^{\alpha }}J_{kj}\frac{%
\delta \varphi ^{j}}{\delta \xi ^{\mu }}=\delta _{k}^{i}.  \label{iii}
\end{equation}
Logo, 
\begin{equation}
J^{\mu \alpha }\frac{\delta \varphi ^{i}}{\delta \xi ^{\alpha }}J_{ij}\frac{%
\delta \varphi ^{j}}{\delta \xi ^{\mu }}=\delta _{i}^{i}=M.  \label{iv}
\end{equation}

Podemos ent\~{a}o computar o tra\c{c}o $\Lambda ^{\mu \mu }$ em (\ref{12}).
Usando que $\delta _{\mu }^{\mu }=2N$ e mais o resultado em (\ref{iv}),
encontramos que 
\begin{equation}
Tr\Lambda =2N-M.  \label{v}
\end{equation}
Nota-se que este coincide com o n\'{u}mero de graus de liberdade do sistema,
ou seja, partindo de um sistema com $2N$ coordenadas simpl\'{e}ticas sujeitas
a $2M$ v\'{\i}nculos de segunda-classe, este ser\'{a} descrito por $2(N-M)$
coordenadas independentes, auto-vetores com auto-valores n\~{a}o-nulos da
matriz do projetor.

Um outro resultado que obtivemos recentemente est\'{a} relacionado com uma
forma alternativa de escrever e ver a estrutura do MPS. Para isso,
escrevamos simplesmente o projetor como 
\begin{equation}
\Lambda =\frac{1}{2}\left( \mathbf{1}+S\right)  \label{vi}
\end{equation}
Naturalmente, o projetor suplementar, que chamaremos V, pode ser escrito
como 
\begin{equation}
V=\frac{1}{2}\left( \mathbf{1}-S\right)  \label{vii}
\end{equation}

Qual seria a motiva\c{c}\~{a}o para escrever estes projetores nesta forma?
Podemos ver facilmente que os auto-vetores da matriz S s\~{a}o as
vari\'{a}veis f\'{\i}sicas, $\xi ^{*},$ com auto-valores $+1$, e as
vari\'{a}veis n\~{a}o-f\'{\i}sicas, $\stackrel{\_}{\xi }^{*}$, com
auto-valores $-1$: 
\begin{equation}
S\xi ^{*}=\left( 2\Lambda -\mathbf{1}\right) \xi ^{*}=\xi ^{*}  \label{viii}
\end{equation}
\begin{equation}
S\stackrel{\_}{\xi }^{*}=-\left( 2V-\mathbf{1}\right) \stackrel{\_}{\xi }%
^{*}=-\stackrel{\_}{\xi }^{*}  \label{ix}
\end{equation}
Isto \'{e}, encontrar os
auto-vetores da matriz S corresponde a encontrar, expl\'{\i}citamente, quais
s\~{a}o as vari\'{a}veis f\'{\i}sicas e quais s\~{a}o as
n\~{a}o-f\'{\i}sicas! \'{E} interessante notar que a condi\c{c}\~{a}o $\det S=+1$ garante a
dimens\~{a}o par para o espa\c{c}o f\'{\i}sico! 

Realmente, encontrar as vari\'{a}veis n\~{a}o-f\'{\i}sicas n\~{a}o \'{e}
algo que usualmente se pretenda. Mas observamos que quando encontramos o
vetor $\mathbf{\xi }^{\ast }$ atrav\'{e}s de (\ref{13}) n\~{a}o encontramos,
em geral, explicitamente as vari\'{a}veis f\'{\i}sicas: algumas das
componentes deste vetor s\~{a}o identicamente nulas, mas algumas s\~{a}o
linearmente dependentes das outras, e enxergar esta depend\^{e}ncia pode
n\~{a}o ser simples em alguns problemas, seja por uma escolha \textit{%
infeliz }de condi\c{c}\~{o}es de \textit{gauge fixing}, seja pela
pr\'{o}pria complexidade do modelo. Ent\~{a}o, pelo menos nestas situa\c{c}%
\~{o}es mais complicadas, a matriz S pode ser a mais \'{u}til.

Tamb\'{e}m vale notar que as observ\'{a}veis da teoria dependem localmente
das coordenadas $\xi ^{\ast }$ sobre a superf\'{\i}cie dos v\'{\i}nculos.
Realmente, como para a Lagrangeana, as condi\c{c}\~{o}es de contorno (\ref{9}%
)(observada a extens\~{a}o natural para o espa\c{c}o de fase) devem ser
tomadas em conta ao se determinar qualquer observ\'{a}vel, o que ultimamente
expressa a independ\^{e}ncia da fun\c{c}\~{a}o correspondente nas
coordenadas locais normais \`{a} superf\'{\i}cie f\'{\i}sica.

Finalmente, uma quest\~{a}o central em teorias de campo com simetria de
gauge, no que diz respeito ao m\'{e}todo de projetores simpl\'{e}ticos,
merece alguns coment\'{a}rios. Como vimos, o MPS toma como ponto de partida
um ``bom'' conjunto de v\'{\i}nculos de segunda classe. Entretanto, a
simetria de gauge em uma teoria de campo implica na exist\^{e}ncia de
v\'{\i}nculos de primeira-classe. Tem-se ent\~{a}o como central a
quest\~{a}o do gauge-fixing. Uma escolha de condi\c{c}\~{o}es de gauge que
``fixe'' o gauge de maneira correta, e que n\~{a}o gere ambiguidades, como
as de Gribov, \'{e} uma condi\c{c}\~{a}o determinante na aplicabilidade do
MPS. Embora existam teorias em que esta escolha n\~{a}o possa ser feita
globalmente, sabemos que localmente esta \'{e} sempre poss\'{\i}vel. Por
outro lado, em todos os casos que aplicamos o MPS at\'{e} o momento, como os
que ser\~{a}o mostrados adiante, tal escolha foi poss\'{\i}vel sem
ambiguidades. Casos em que esta escolha s\'{o} \'{e} possivel localmente
dever\~{a}o merecer um estudo futuro.

\chapter{{\sc Modelos Ilustrativos}}

Com o objetivo de ilustrar, em sistemas simples e did\'{a}ticos, o MPS,
analisaremos dois exemplos de teorias n\~{a}o-relativ\'{\i}sticas que
v\^{e}m sendo discutidas na literatura.

\section{{\sc Modelo de Christ-Lee}}

Um modelo mec\^{a}nico muito simples, n\~{a}o-relativ\'{\i}stico, invariante
de gauge, foi utilizado originalmente por N.H. Christ e T.D. Lee\cite
{christ-lee} para ilustrar consequ\^{e}ncias de diferentes escolhas de gauge
em uma teoria. Desde ent\~{a}o, este tem sido utilizado por v\'{a}rios
autores, com prop\'{o}sito de exemplificar diferentes m\'{e}todos de quantiza%
\c{c}\~{a}o (\cite{costa-girotti},\cite{phokhorov},\cite{bouzas}). Em uma
destas an\'{a}lises \cite{costa-girotti} o espa\c{c}o de fase
``f\'{\i}sico'' foi obtido, embora atrav\'{e}s de argumentos intuitivos.
Mostramos ent\~{a}o \cite{c-lee} como o mesmo resultado pode ser obtido de
maneira sistem\'{a}tica atrav\'{e}s do MPS.

A Lagrangeana de partida \'{e} 
\begin{eqnarray}
L &=&\frac{1}{2}\left( \stackrel{.}{x}_{1}^{2}+\stackrel{.}{x}%
_{2}^{2}\right) -\left( x_{1}\stackrel{.}{x}_{2}-x_{2}\stackrel{.}{x}%
_{1}\right) \\
&&+\frac{1}{2}x_{3}^{2}\left( x_{1}^{2}+x_{2}^{2}\right)
-V(x_{1}^{2}+x_{2}^{2})  \label{2.1}
\end{eqnarray}
A Hamiltoniana can\^{o}nica associada e os respectivos v\'{\i}nculos de
segunda-classe (ap\'{o}s o gauge-fixing) s\~{a}o: 
\begin{equation}
H=\frac{1}{2}p_{1}^{2}+\frac{1}{2}p_{2}^{2}+V(x_{1}^{2}+x_{2}^{2})
\label{2.2}
\end{equation}
e 
\begin{eqnarray}
\varphi _{1} &=&p_{3}=0  \label{2.2a} \\
\varphi _{2} &=&p_{2}-ep_{1}=0  \label{2.2b} \\
\varphi _{3} &=&x_{2}-ex_{1}=0  \label{2.2c} \\
\varphi _{4} &=&x_{3}=0  \label{2.2d}
\end{eqnarray}
onde $e=\tan b/c$ , e $b$ e $c$ s\~{a}o constantes n\~{a}o-nulas \cite
{costa-girotti}. Observamos que a contagem de graus de liberdade j\'{a} nos
permite esperar que o espa\c{c}o de fase f\'{\i}sico ter\'{a} $6-4=2$
dimens\~{o}es, o que dever\'{a} ser confirmado pelo MPS.

As vari\'{a}veis simpl\'{e}ticas podem ser definidas atrav\'{e}s da
correspond\^{e}ncia: 
\begin{equation}
(x_{1},x_{2},x_{3},p_{1},p_{2},p_{3})\leftrightarrow (\xi _{1},\xi _{2},\xi
_{3},\xi _{4},\xi _{5},\xi _{6})  \label{2.3}
\end{equation}
O projetor simpl\'{e}tico assume, neste caso, a forma simplificada (\ref{8}%
), onde a m\'{e}trica local (\ref{5}) pode ser facilmente calculada tendo em
conta (\ref{2.2a}-\ref{2.2d}); temos, explicitamente: 
\begin{equation}
g=\left( 
\begin{array}{cccc}
0 & 0 & 0 & -1 \\ 
0 & 0 & -(1+e^{2}) & 0 \\ 
0 & (1+e^{2}) & 0 & 0 \\ 
1 & 0 & 0 & 0
\end{array}
\right) ,  \label{2.4}
\end{equation}
\begin{equation}
g^{-1}=\left( 
\begin{array}{cccc}
0 & 0 & 0 & 1 \\ 
0 & 0 & \left( 1+e^{2}\right) ^{-1} & 0 \\ 
0 & -\left( 1+e^{2}\right) ^{-1} & 0 & 0 \\ 
-1 & 0 & 0 & 0
\end{array}
\right) ,  \label{2.5}
\end{equation}
e 
\begin{equation}
\Lambda =\left( 
\begin{array}{cccccc}
\left( 1+e^{2}\right) ^{-1} & e\left( 1+e^{2}\right) ^{-1} & 0 & 0 & 0 & 0
\\ 
e\left( 1+e^{2}\right) ^{-1} & e^{2}\left( 1+e^{2}\right) ^{-1} & 0 & 0 & 0
& 0 \\ 
0 & 0 & 0 & 0 & 0 & 0 \\ 
0 & 0 & 0 & \left( 1+e^{2}\right) ^{-1} & e\left( 1+e^{2}\right) ^{-1} & 0
\\ 
0 & 0 & 0 & e\left( 1+e^{2}\right) ^{-1} & e^{2}\left( 1+e^{2}\right) ^{-1}
& 0 \\ 
0 & 0 & 0 & 0 & 0 & 0
\end{array}
\right) \;. \label{2.6}
\end{equation}
O tra\c{c}o desta matriz nos d\'{a} a primeira confirma\c{c}\~{a}o a
respeito da contagem dos graus de liberdade do sistema: $Tr\Lambda =2.$

Com esta matriz em (\ref{13}) obtemos as vari\'{a}veis projetadas: 
\begin{eqnarray}
\xi _{1}^{\ast } &=&\left( 1+e^{2}\right) ^{-1}\xi _{1}+e\left(
1+e^{2}\right) ^{-1}\xi _{2}  \label{2.7a} \\
\xi _{2}^{\ast } &=&e\xi _{1}^{\ast }  \label{2.7b} \\
\xi _{3}^{\ast } &=&0  \label{2.7c} \\
\xi _{4}^{\ast } &=&\left( 1+e^{2}\right) ^{-1}\xi _{4}+e\left(
1+e^{2}\right) ^{-1}\xi _{5}  \label{2.7d} \\
\xi _{5}^{\ast } &=&e\xi _{4}^{\ast }  \label{2.7e} \\
\xi _{6}^{\ast } &=&0  \label{2.7f}
\end{eqnarray}
Vemos ent\~{a}o, explicitamente, que o conjunto de vari\'{a}veis projetadas
possui apenas duas vari\'{a}veis independentes, $\xi _{1}^{*}$ e $\xi
_{4}^{*} $, que s\~{a}o as vari\'{a}veis f\'{\i}sicas do problema.

Para completar a descri\c{c}\~{a}o deste modelo vamos encontrar as equa\c{c}%
\~{o}es do movimento neste espa\c{c}o reduzido. A Hamiltoniana (\ref{2.2})
na forma simpl\'{e}tica \'{e} 
\begin{equation}
H=\frac{1}{2}\xi _{4}^{2}+\frac{1}{2}\xi _{5}^{2}+V(\xi _{1}^{2}+\xi\;.
_{2}^{2})  \label{2.8}
\end{equation}
Ap\'{o}s projetada, temos a Hamiltoniana f\'{\i}sica 
\begin{eqnarray}
H^{*} &=&\frac{1}{2}\xi _{4}^{*2}+\frac{1}{2}\xi _{5}^{*2}+V(\xi
_{1}^{*2}+\xi _{2}^{*2}) \\
&=&\frac{1}{2}\left( 1+e^{2}\right) \xi _{4}^{*2}+V(\left( 1+e^{2}\right)
\xi _{1}^{*2}) \;. \label{2.9}
\end{eqnarray}
Usando novas vari\'{a}veis 
\begin{eqnarray}
x_{*} &=&\left( 1+e^{2}\right) ^{1/2}\xi _{1}^{*}  \label{2.10a} \\
p_{*} &=&\left( 1+e^{2}\right) ^{1/2}\xi _{4}^{*}  \label{2.10b}
\end{eqnarray}
a Hamiltoniana f\'{\i}sica assume a forma mais familiar 
\begin{equation}
H^{*}=\frac{1}{2}p_{*}^{2}+V(x_{*}^{2})  \label{2.11}
\end{equation}
As equa\c{c}\~{o}es can\^{o}nicas do movimento s\~{a}o ent\~{a}o obtidas
atrav\'{e}s das equa\c{c}\~{o}es de Hamilton-Jacobi usuais: 
\begin{eqnarray}
\stackrel{.}{x}_{*} &=&\{x_{*},H^{*}\}=p_{*}  \label{2.12a} \\
\stackrel{.}{p}_{*} &=&\{p_{*},H^{*}\}=-\frac{\partial V}{\partial x_{*}}\;
\label{2.12b}
\end{eqnarray}
Estes resultados concordam com aqueles obtidos em \cite{costa-girotti}
usando um formalismo completamente independente.

\section{{\sc Part\'{\i}cula em um Campo Eletromagn\'etico}}

No exemplo anterior o MPS foi aplicado em um modelo com simetria de gauge
utilizando apenas a abordagem geom\'{e}trica. A \'{u}nica refer\^{e}ncia aos
m\'{e}todos usuais de formula\c{c}\~{a}o can\^{o}nica de teorias de gauge
encontra-se na maneira de analisar a estrutura dos v\'{\i}nculos presentes
no modelo a fim de determinar qual o conjunto m\'{\i}nimo de v\'{\i}nculos
de segunda-classe existente. Entretanto, como visto anteriormente (\ref{m11}%
), \'{e} poss\'{\i}vel tratar o problema de forma mais aproximada aos
m\'{e}todos tradicionais, utilizando os par\^{e}nteses de Dirac fundamentais
como ponto de partida na constru\c{c}\~{a}o dos projetores. Mostraremos
agora um exemplo simples em que esta abordagem \'{e} utilizada.

O seguinte ``toy model'' tem sido utilizado recentemente (\cite{bigatti}, 
\cite{kim}) em discuss\~{o}es a respeito das propriedades
n\~{a}o-comutativas de cordas abertas e D-branas na presen\c{c}a de um campo
B \cite{mmi}. O ponto de partida \'{e} a seguinte Lagrangeana 
\begin{equation}
L=-\frac{eB}{2c}\dot{x}^{i}\varepsilon _{ij}\,x^{j}+e\Phi (x),  \label{2.13}
\end{equation}
que descreve uma part\'{\i}cula em intera\c{c}\~{a}o com um campo
eletro-magn\'{e}tico com potencial est\'{a}tico $\Phi (x)$ e campo
magn\'{e}tico constante e intenso B, no limite onde o termo de energia
cin\'{e}tica pode ser desprezado ou $m\rightarrow 0$. O espa\c{c}o de fase
do modelo \'{e} 4-dimensional e suas coordenadas est\~{a}o sujeitas a dois
v\'{\i}nculos de segunda-classe 
\begin{equation}
\varphi _{i}\equiv p_{i}+\frac{eB}{2c}\;\varepsilon _{ij}\,x^{j}\approx 0,
\label{2.14}
\end{equation}
o que nos leva a uma contagem de $4-2=2$ graus de liberdade.

Os par\^{e}nteses de Dirac (\ref{m10}) das vari\'{a}veis originais s\~{a}o
os seguintes: 
\begin{eqnarray}
\{x^{i}\;,\;x^{j}\}_{D}&=&\frac{c}{eB}\,\varepsilon ^{ij}  \label{2.15} \\
\{x^{i}\;,\;p_{j}\}_{D}&=&\frac{1}{2}\delta _{j}^{\,i}  \label{2.16} \\
\{p_{i}\;,\;p_{j}\}_{D}&=&-\frac{eB}{4c}\,\varepsilon _{ij}\;.  \label{2.17}
\end{eqnarray}
Vamos introduzir o vetor simpl\'{e}tico $\mathbf{\xi }$ com as seguintes
componentes 
\begin{equation}
\left( \xi ^{1},\xi ^{2},\xi ^{3},\xi ^{4}\right) \equiv \left(
x^{1},x^{2},p_{1},p_{2}\right) .  \label{2.18}
\end{equation}
\'{E} f\'{a}cil ver que a matriz D \'{e} 
\begin{equation}
D={\left( 
\begin{array}{cccc}
0 & -\frac{c}{eB} & \frac{1}{2} & 0 \\ 
\frac{c}{eB} & 0 & 0 & \frac{1}{2} \\ 
-\frac{1}{2} & 0 & 0 & -\frac{eB}{4c} \\ 
0 & -\frac{1}{2} & \frac{eB}{4c} & 0
\end{array}
\right)} \;.  \label{2.19}
\end{equation}
Usando a rela\c{c}\~{a}o (\ref{m11}) podemos escrever o projetor
simpl\'{e}tico 
\begin{equation}
\Lambda ={\left( 
\begin{array}{cccc}
\frac{1}{2} & 0 & 0 & \frac{c}{eB} \\ 
0 & \frac{1}{2} & -\frac{c}{eB} & 0 \\ 
0 & -\frac{eB}{4c} & \frac{1}{2} & 0 \\ 
\frac{eB}{4c} & 0 & 0 & \frac{1}{2}
\end{array}
\right) }\;.  \label{2.20}
\end{equation}
(Observe-se que o tra\c{c}o desta matriz reproduz a contagem dos graus de
liberdade feita acima.)

Com (\ref{2.20}) encontramos as vari\'{a}veis projetadas: 
\begin{eqnarray}
\xi ^{\ast 1}&=&\frac{1}{2}\,\xi ^{1}+\frac{c}{eB}\,\xi ^{4}  \label{2.21} \\
\xi ^{\ast 2}&=&\frac{1}{2}\,\xi ^{2}-\frac{c}{eB}\,\xi ^{3}  \label{2.22}\\
\xi ^{\ast 3}&=&-\frac{eB}{4c}\,\xi ^{2}+\frac{1}{2}\,\xi ^{3}=-\frac{eB}{2c}%
\,\xi ^{\ast 2}  \label{2.23} \\
\xi ^{\ast 4}&=&\frac{eB}{4c}\,\xi ^{1}+\frac{1}{2}\,\xi ^{4}=\frac{eB}{2c}%
\,\xi ^{\ast 1}\;.  \label{2.24}
\end{eqnarray}
Vemos ent\~{a}o que o espa\c{c}o de fase reduzido \'{e} bi-dimensional com
as rela\c{c}\~{o}es can\^{o}nicas de comuta\c{c}\~{a}o usuais entre $\xi
^{*1}$ e $\xi ^{*2}$ . A Hamiltoniana expressa em termos destas duas
vari\'{a}veis f\'{\i}sicas pode ser usada como ponto de partida para a
quantiza\c{c}\~{a}o do sistema.

\chapter{{\sc Aplica\c c\~oes em Teorias de Campos de Gauge}}

Neste cap\'{\i}tulo vamos aplicar o m\'{e}todo dos projetores
simpl\'{e}ticos a teorias de campos. Teorias de gauge em 2, 3 e 4
dimens\~{o}es s\~{a}o analisadas com o intuito de testar a aplicabilidade do
m\'{e}todo, assim como ilustrar sua efic\'{a}cia.

\section{{\sc Eletrodin\^amica 4-$D$}}

Como primeiro exemplo de aplica\c{c}\~{a}o do MPS em teorias de campo vamos
analisar o mo\-de\-lo cl\'{a}ssico da eletrodin\^{a}mica de Maxwell no gauge de
radia\c{c}\~{a}o, uma vez que a Hamiltoniana em termos dos campos
f\'{\i}sicos para este modelo \'{e} um objeto bastante conhecido \cite
{sakurai}.

Como ponto de partida tomaremos a Hamiltoniana can\^{o}nica 
\begin{equation}
\mathcal{H}_c=\int d^{3}x\left\{ \frac{1}{2}(\boldmath\mbox{$\pi$}^{2}+\mathbf{B}%
^{2})+\pi _{\psi }\gamma ^{0}(\mbox{\boldmath$\gamma \cdot \partial$}-ie\,\mbox{\boldmath$\gamma
\cdot A$}-im)\psi \right\}
\label{301}
\end{equation}
sujeita ao conjunto de v\'{\i}nculos de segunda classe 
\begin{eqnarray}
\varphi _{1} &=&\pi _{0}  \label{302.a} \\
\varphi _{2} &=& \mbox{\boldmath$\partial \cdot \pi$}+ie\;\pi _{\psi }\psi \label{302.b} \\
\varphi _{3} &=&A_{0}  \label{302.c} \\
\varphi _{4} &=& \mbox{\boldmath$\partial \cdot A$} \label{302.d}
\end{eqnarray}
Estas s\~{a}o as informa\c{c}\~{o}es b\'{a}sicas das quais necessitamos para
aplicar o MPS: a m\'{e}trica local $J_{ij}\left( x,y\right) $, \'{e} dada
pela matriz 
\begin{equation}
\left( 
\begin{array}{cccc}
0 & 0 & \delta ^{3}(x-y) & 0 \\ 
0 & 0 & 0 & \nabla ^{-2} \\ 
-\delta ^{3}(x-y) & 0 & 0 & 0 \\ 
0 & -\nabla ^{-2} & 0 & 0
\end{array}
\right) ,  \label{303}
\end{equation}
e com a prescri\c{c}\~{a}o 
\begin{equation}
\left( A_{0},A_{1},A_{2},A_{3},\psi ,\pi _{0},\pi _{1},\pi _{2},\pi _{3},\pi
_{\psi }\right) \longleftrightarrow \left( \xi _{1},\cdots,\xi _{10}\right)
\label{304}
\end{equation}
podemos calcular, atrav\'{e}s de (\ref{12}), a matriz do projetor
simpl\'{e}tico. Uma simples \'{a}lgebra fornece  a matriz $\Lambda(x,y)$
\begin{eqnarray}
&&{\hspace{-1cm}\left( 
\begin{array}{cccccccccc}
0 \! \!&\! \! 0 \! \!&\! \! 0 \! \!&\! \! 0 \! \!&\! \! 0 \! \!&\! \! 0 \! \!&\! \! 0
\! \!&\! \! 0 \! \!&\! \! 0 \! \!&\! \! 0 \\ 
0 \! \!&\! \! {\scriptstyle\delta ^{3}(x-y)-}\frac{\partial _{1}^{x}\partial
_{1}^{y}}{\nabla ^{2}} \! \!&\! \! -\frac{\partial _{1}^{x}\partial _{2}^{y}}{%
\nabla ^{2}} \! \!&\! \! -\frac{\partial _{1}^{x}\partial _{3}^{y}}{\nabla ^{2}%
} \! \!&\! \! 0 \! \!&\! \! 0 \! \!&\! \! 0 \! \!&\! \! 0 \! \!&\! \! 0 \! \!&\! \! 0 \\ 
0 \! \!&\! \! -\frac{\partial _{2}^{x}\partial _{1}^{y}}{\nabla ^{2}} \! \!&\! \! 
{\scriptstyle\delta ^{3}(x-y)-}\frac{\partial _{2}^{x}\partial _{2}^{y}}{%
\nabla ^{2}} \! \!&\! \! -\frac{\partial _{2}^{x}\partial _{3}^{y}}{\nabla ^{2}%
} \! \!&\! \! 0 \! \!&\! \! 0 \! \!&\! \! 0 \! \!&\! \! 0 \! \!&\! \! 0 \! \!&\! \! 0 \\ 
0 \! \!&\! \! -\frac{\partial _{3}^{x}\partial _{1}^{y}}{\nabla ^{2}} \! \!&\! \! 
-\frac{\partial _{3}^{x}\partial _{2}^{y}}{\nabla ^{2}} \! \!&\! \! {%
\scriptstyle\delta ^{3}(x-y)-}\frac{\partial _{3}^{x}\partial _{3}^{y}}{%
\nabla ^{2}} \! \!&\! \! 0 \! \!&\! \! 0 \! \!&\! \! 0 \! \!&\! \! 0 \! \!&\! \! 0
\! \!&\! \! 0 \\ 
0 \! \!&\! \! \frac{-ie\,\xi _{5}(x)\partial _{1}^{y}}{\nabla ^{2}} \! \!&\! \! %
\frac{-ie\,\xi _{5}(x)\partial _{2}^{y}}{\nabla ^{2}} \! \!&\! \! \frac{%
-ie\,\xi _{5}(x)\partial _{3}^{y}}{\nabla ^{2}} \! \!&\! \! {\scriptstyle%
\delta ^{3}(x-y)} \! \!&\! \! 0 \! \!&\! \! 0 \! \!&\! \! 0 \! \!&\! \! 0 \! \!&\! \! 0
\\ 
0 \! \!&\! \! 0 \! \!&\! \! 0 \! \!&\! \! 0 \! \!&\! \! 0 \! \!&\! \! 0 \! \!&\! \! 0
\! \!&\! \! 0 \! \!&\! \! 0 \! \!&\! \! 0 \\ 
0 \! \!&\! \! 0 \! \!&\! \! 0 \! \!&\! \! 0 \! \!&\! \! \frac{ie\partial _{1}^{x}\xi
_{10}(y)}{\nabla ^{2}} \! \!&\! \! 0 \! \!&\! \! {\scriptstyle\delta ^{3}(x-y)-}%
\frac{\partial _{1}^{x}\partial _{1}^{y}}{\nabla ^{2}} \! \!&\! \! -\frac{%
\partial _{1}^{x}\partial _{2}^{y}}{\nabla ^{2}} \! \!&\! \! -\frac{\partial
_{1}^{x}\partial _{3}^{y}}{\nabla ^{2}} \! \!&\! \! \frac{-ie\partial
_{1}^{x}\xi _{5}(y)}{\nabla ^{2}} \\ 
0 \! \!&\! \! 0 \! \!&\! \! 0 \! \!&\! \! 0 \! \!&\! \! \frac{ie\partial _{2}^{x}\xi
_{10}(y)}{\nabla ^{2}} \! \!&\! \! 0 \! \!&\! \! -\frac{\partial
_{2}^{x}\partial _{1}^{y}}{\nabla ^{2}} \! \!&\! \! {\scriptstyle\delta
^{3}(x-y)-}\frac{\partial _{2}^{x}\partial _{2}^{y}}{\nabla ^{2}} \! \!&\! \! -%
\frac{\partial _{2}^{x}\partial _{3}^{y}}{\nabla ^{2}} \! \!&\! \! \frac{%
-ie\partial _{2}^{x}\xi _{5}(y)}{\nabla ^{2}} \\ 
0 \! \!&\! \! 0 \! \!&\! \! 0 \! \!&\! \! 0 \! \!&\! \! \frac{ie\partial _{3}^{x}\xi
_{10}(y)}{\nabla ^{2}} \! \!&\! \! 0 \! \!&\! \! -\frac{\partial
_{3}^{x}\partial _{1}^{y}}{\nabla ^{2}} \! \!&\! \! -\frac{\partial
_{3}^{x}\partial _{2}^{y}}{\nabla ^{2}} \! \!&\! \! {\scriptstyle\delta
^{3}(x-y)-}\frac{\partial _{3}^{x}\partial _{3}^{y}}{\nabla ^{2}} \! \!&\! \! %
\frac{-ie\partial _{3}^{x}\xi _{5}(y)}{\nabla ^{2}} \\ 
0 \! \!&\! \! \frac{ie\,\xi _{10}(x)\partial _{1}^{y}}{\nabla ^{2}} \! \!&\! \! %
\frac{ie\,\xi _{10}(x)\partial _{2}^{y}}{\nabla ^{2}} \! \!&\! \! \frac{%
ie\,\xi _{10}(x)\partial _{3}^{y}}{\nabla ^{2}} \! \!&\! \! 0 \! \!&\! \! 0
\! \!&\! \! 0 \! \!&\! \! 0 \! \!&\! \! 0 \! \!&\! \! {\scriptstyle\delta ^{3}(x-y)}
\end{array}
\right)}  \label{305} \nonumber \\
\end{eqnarray}
que ao ser aplicada ao vetor simpl\'{e}tico produz as coordenadas
f\'{\i}sicas 
\begin{eqnarray}
\xi ^{1*}\left( x\right) &=& \xi ^{6*}\left( x\right) =0  \label{305.a} \\
\xi ^{n*}\left( x\right) &=& A_{n-1}^{\perp }\left( x\right) ,n=2,3,4
\label{305.b}\\
\xi ^{5*}\left( x\right) &=&\psi \left( x\right)  \label{305.c} \\
\xi ^{10*}\left( x\right) &=&\pi _{\psi }\left( x\right)  \label{305.d} \\
\xi ^{m*}(x)&=&\pi _{m-6}^{\perp }(x)+2ie\int d^{\,3}y\,\partial
_{m-6}^{x}\nabla ^{-2}\pi _{\psi }(y)\,\psi (y)\;,m=7,8,9\;.  \label{305.e}
\end{eqnarray}

Para encontrar a Hamiltoniana projetada reescrevemos a Hamiltoniana
vinculada (\ref{301}) na nota\c{c}\~{a}o simpl\'{e}tica: 
\begin{eqnarray}
\mathcal{H} &=&\int d^{3}x \{\frac{1}{2}(\xi _{7}^{2}+\xi _{8}^{2}+\xi
_{9}^{2})+\frac{1}{2}(\varepsilon _{\alpha \beta \gamma }\partial _{\beta
}\xi _{\gamma })^{2}  \nonumber \\
&&+ \xi _{10}\gamma ^{0}[\mbox{\boldmath$\gamma \cdot \partial$} -ie(\gamma _{1}\xi
_{2}+\gamma _{2}\xi _{3}+\gamma _{3}\xi _{4})-im]\xi _{5} \},  \label{306}
\end{eqnarray}
com $\alpha ,\beta ,\gamma =2,3,4.$ A Hamiltoniana f\'{\i}sica \'{e} ent\~{a}o: 
\begin{eqnarray}
\mathcal{H}^{*} &=&\int d^{3}x\{\frac{1}{2}(\xi _{7}^{*2}+\xi _{8}^{*2}+\xi
_{9}^{*2})+\frac{1}{2}(\varepsilon _{\alpha \beta \gamma }\partial _{\beta
}\xi _{\gamma }^{*})^{2}  \nonumber \\
&&+\xi _{10}^{*}\gamma ^{0}[\mbox{\boldmath$\gamma \cdot \partial$}-ie(\gamma _{1}\xi
_{2}^{*}+\gamma _{2}\xi _{3}^{*}+\gamma _{3}\xi _{4}^{*})-im]\xi _{5}^{*}\}.
\label{307}
\end{eqnarray}
Retornando \`{a} nota\c{c}\~{a}o do espa\c{c}o de fase original,
atrav\'{e}s de (\ref{305.a}-\ref{305.e}), temos finalmente: 
\begin{eqnarray}
\mathcal{H}^{*}&=&\int d^{\,3}x\,\frac{1}{2}\left(\mbox{\boldmath$\pi$}^{\perp 2}+
\mbox{\boldmath$B$}^{2}\right)
+\pi _{\psi}\gamma^{0}\left(\mbox{\boldmath$\gamma\cdot\partial$}-ie\,\mbox{%
\boldmath$\gamma\cdot{A}$}^{\perp}-im\right)\psi \nonumber \\
&&+\frac{e^{2}}{2\pi}\int d^{\,3}x\,d^{\,3}y~\pi _{\psi
}(x)\,\psi(x)~\frac{1}{4\pi|\mbox{\boldmath$\,x-y\,$}|}\pi _{\psi}(y)\,\psi(y)\;.  
\label{308}
\end{eqnarray}
que \'{e} a forma familiar da Hamiltoniana de Fermi.

\'{E} interessante notar que, conforme observamos ao final do cap\'{\i}tulo
anterior, embora o tra\c{c}o da matriz do projetor forne\c{c}a a
dimens\~{a}o exata do espa\c{c}o f\'{\i}sico, as vari\'{a}veis projetadas,
neste caso, n\~{a}o s\~{a}o exatamente aquelas que representam este espa\c{c}%
o, sendo necess\'{a}rio escolher uma determinada dire\c{c}\~{a}o de propaga%
\c{c}\~{a}o \textit{a posteriori}. Neste contexto, talvez fosse interessante
pesquisar os auto-vetores da matriz S, ou, ainda, fazer uma escolha de gauge
mais \textit{adequada.} Como nosso intuito aqui \'{e} a compara\c{c}\~{a}o
com resultados j\'{a} estabelecidos, nos damos por satisfeitos com esta
ilustra\c{c}\~{a}o.

\section{{\sc Modelo Bosonizado de Schwinger 2-$D$}}

Como ilustra\c{c}\~{a}o do MPS no caso de uma teoria de gauge
bi-dimensional, analisaremos um modelo, proposto por Schwinger (\cite
{schwinger},\cite{swieca}), dado pela densidade Lagrangiana 
\begin{equation}
\mathcal{L=}\stackrel{\_}{\psi }\gamma ^{\mu }(i\partial _{\mu }-eA_{\mu
})\psi -\frac{1}{4}F_{\mu \nu }F^{\mu \nu }.  \label{309}
\end{equation}
Usando o dicion\'{a}rio de bosoniza\c{c}\~{a}o de Kogut e Susskind \cite
{kogut}, 
\begin{equation}
i\stackrel{\_}{\psi }\gamma ^{\mu }\partial _{\mu }\psi =\frac{1}{2}\left(
\partial _{\mu }\phi \right) \left( \partial ^{\mu }\phi \right)  \label{310}
\end{equation}
\begin{equation}
\stackrel{\_}{\psi }\gamma ^{\mu }\psi =\frac{1}{\sqrt{\pi }}\varepsilon
^{\mu \nu }\partial _{\nu }\phi ,  \label{311}
\end{equation}
obtemos a vers\~{a}o bosonizada deste modelo como: 
\begin{equation}
\mathcal{L=}\frac{1}{2}\,\partial _{\mu }\phi \,\partial ^{\mu }\phi
+e\,\varepsilon ^{\mu \nu }\,\partial _{\mu }\phi A_{\nu }-\frac{1}{4}F_{\mu \nu
}F^{\mu \nu }.  \label{312}
\end{equation}
Aplicaremos o m\'{e}todo dos projetores simpl\'{e}ticos nesta teoria
bosonizada.

Os momenta can\^{o}nicos obtidos a partir de (\ref{312}) s\~{a}o: 
\begin{equation}
\pi _{0}\approx 0  \label{313.1}
\end{equation}
\begin{equation}
\pi _{1}=F_{10}=\partial _{1}A_{0}-\partial _{0}A_{1}  \label{313.2}
\end{equation}
\begin{equation}
\pi _{\phi }=\partial _{0}\phi +eA_{1}.  \label{313.3}
\end{equation}
A Hamiltoniana prim\'{a}ria \'{e}: 
\begin{equation}
H =\int dx\left[\frac{1}{2}\left\{\pi _{1}^{2}+\pi _{\phi }^{2}+(\partial _{1}\phi
)^{2}-e^{2}A_{1}^{2}\right\} +e\pi _{\phi }A_{1}+A_{0}(\partial _{1}\pi _{1}+e\partial _{1}\phi
)+\lambda \pi _{0}\right]
\end{equation}
onde $\lambda $ \'{e} um campo multiplicador de Lagrange.

Impondo condi\c{c}\~{o}es de consist\^{e}ncia sobre o v\'{\i}nculo
prim\'{a}rio (\ref{313.1}), obtemos apenas o v\'{\i}nculo secund\'{a}rio 
\begin{equation}
\partial _{1}\pi _{1}+e\partial _{1}\phi \approx 0.  \label{315}
\end{equation}

A estes v\'{\i}nculos de primeira-classe adicionamos as condi\c{c}\~{o}es de
``Gauge de Coulomb'', de forma a termos o seguinte conjunto de v\'{\i}nculos
de segunda-classe na teoria: 
\begin{equation}
\varphi _{1}=\pi _{0}=0  \label{316.a}
\end{equation}
\begin{equation}
\varphi _{2}=\partial _{1}\pi _{1}+e\partial _{1}\phi =0  \label{316.b}
\end{equation}
\begin{equation}
\varphi _{3}=A_{0}=0  \label{316.c}
\end{equation}
\begin{equation}
\varphi _{4}=\partial _{1}A_{1}=0.  \label{316.d}
\end{equation}
Com este conjunto construimos a matriz $g_{ij}\left( x,y\right) =\{\varphi
_{i}(x),\varphi _{j}(y)\}:$%
\begin{equation}
g=\left( 
\begin{array}{cccc}
0 & 0 & -\delta (x-y) & 0 \\ 
0 & 0 & 0 & -\partial _{1}^{2}\delta (x-y) \\ 
\delta (x-y) & 0 & 0 & 0 \\ 
0 & \partial _{1}^{2}\delta (x-y) & 0 & 0
\end{array}
\right) ,  \label{317}
\end{equation}
tendo como inversa 
\begin{equation}
g^{-1}=\left( 
\begin{array}{cccc}
0 & 0 & \delta (x-y) & 0 \\ 
0 & 0 & 0 & \frac{1}{\partial _{1}^{2}} \\ 
-\delta (x-y) & 0 & 0 & 0 \\ 
0 & -\frac{1}{\partial _{1}^{2}} & 0 & 0
\end{array}
\right) .  \label{318}
\end{equation}

Os elementos de matriz do projetor, $\Lambda _{\nu }^{\mu }\left( x,y\right) 
$, podem ent\~{a}o ser calculados a partir da defini\c{c}\~{a}o geral (\ref
{12}). Usando a prescri\c{c}\~{a}o 
\begin{equation}
\left( A_{0},A_{1},\phi ,\pi _{0},\pi _{1},\pi _{\phi }\right)
\longleftrightarrow \left( \xi _{1},...,\xi _{6}\right) ,  \label{319}
\end{equation}
encontramos a matriz $\Lambda $ na seguinte forma: 
\begin{equation}
\Lambda \left( x,y\right) =\left( 
\begin{array}{cccccc}
0 & 0 & 0 & 0 & 0 & 0 \\ 
0 & \delta (x-y)+\frac{\partial _{1}^{x}\partial _{2}^{y}}{\partial _{1}^{2}}
& 0 & 0 & 0 & 0 \\ 
0 & 0 & \delta (x-y) & 0 & 0 & 0 \\ 
0 & 0 & 0 & 0 & 0 & 0 \\ 
0 & 0 & e\frac{\partial _{1}^{x}\partial _{2}^{y}}{\partial _{1}^{2}} & 0 & 
\delta (x-y)+\frac{\partial _{1}^{x}\partial _{2}^{y}}{\partial _{1}^{2}} & 0
\\ 
0 & e\frac{\partial _{1}^{x}\partial _{2}^{y}}{\partial _{1}^{2}} & 0 & 0 & 0
& \delta (x-y)
\end{array}
\right)  \label{320}
\end{equation}
Assim, as coordenadas projetadas s\~{a}o: 
\begin{eqnarray}
\xi _{1}^{*}\left( x\right) &=&0  \label{321.a} \\
\xi _{2}^{*}\left( x\right) &=&0  \label{321.b} \\
\xi _{3}^{*}\left( x\right) &=&\phi \left( x\right)  \label{321.c} \\
\xi _{4}^{*}\left( x\right) &=&0  \label{321.d} \\
\xi _{5}^{*}\left( x\right) &=&-e\phi \left( x\right)  \label{321.e} \\
\xi _{6}^{*}\left( x\right) &=&-eA_{1}\left( x\right) +\pi _{\phi }\left(
x\right)  \label{321.f}
\end{eqnarray}
Vemos da\'{\i} que o espa\c{c}o f\'{\i}sico em quest\~{a}o possui apenas
duas coordenadas, $\xi _{3}^{*}$ e $\xi _{6}^{*}$, uma vez que $\xi
_{5}^{*}=-e\xi _{3}^{*}.$

A Hamiltoniana vinculada, escrita na nota\c{c}\~{a}o simpl\'{e}tica, \'{e}: 
\begin{equation}
H =\int dx \left\{ \frac{1}{2}\left[\xi _{5}^{2}+\xi _{6}^{2}+(\partial _{1}\xi
_{3})^{2}-e^{2}\xi _{2}^{2}\right] +e\xi _{6}\xi _{2}+\xi _{1}(\partial _{1}\xi _{5}
+e\partial _{1}\xi
_{3})+\lambda \xi _{4}\right\}.  \label{322}
\end{equation}
A Hamiltoniana projetada \'{e} ent\~{a}o: 
\begin{equation}
H^{\ast }=\int dx\frac{1}{2}\left[\xi _{5}^{\ast 2}+\xi _{6}^{\ast 2}+(\partial
_{1}\xi _{3}^{\ast })^{2}\right].  \label{323}
\end{equation}
Das rela\c{c}\~{o}es (\ref{321.a}-\ref{321.f}) notamos que 
\begin{equation}
\xi _{5}^{\ast }=-e\xi _{3}^{\ast },  \label{324}
\end{equation}

o que significa que esta n\~{a}o \'{e} uma variavel independente. Desta
forma, a express\~{a}o (\ref{323}) se torna: 
\begin{equation}
H^{\ast }=\int dx\frac{1}{2}[e^{2}q^{2}+p^{2}+(\partial _{1}q)^{2}],
\label{325}
\end{equation}
onde escrevemos $(q,p)$ no lugar de $(\xi _{3}^{\ast },\xi _{6}^{\ast })$.
Notamos que $p\equiv \pi _{\phi }\left( x\right) -eA_{1}\left( x\right) .$
Este corresponde ao novo campo introduzido \textit{ad hoc }em \cite{clovis}.
Desta Hamiltoniana podemos ver que:
$$\dot{q}=\left\{ q,H^{\ast }\right\} =p$$ 
e 
$$\ddot{q}=\dot{p}=e^{2}q+\partial \partial _{1}^{2}q,$$ 
ou seja, 
\begin{equation}
\left( \square +e^{2}\right) \phi =0.  \label{326}
\end{equation}
Portanto, $\phi $ \'{e} um campo massivo livre com massa igual a $e$.

\section{{\sc Chern-Simons-Maxwell sem Mat\'eria}}

Neste exemplo iremos derivar a Hamiltoniana f\'{\i}sica e as equa\c{c}%
\~{o}es de movimento para o modelo 3D de Chern-Simons-Maxell sem a presen\c{c%
}a de campos de mat\'{e}ria. Usaremos as condi\c{c}\~{o}es de gauge de
Coulomb a fim de construir um conjunto de v\'{\i}nculos de segunda-classe. A
express\~{a}o da Hamiltoniana f\'{\i}sica est\'{a} muito pr\'{o}xima de uma
outra obtida em um trabalho recente\cite{devecchi} onde o procedimento de
quantisa\c{c}\~{a}o via par\^{e}nteses de Dirac (DBQP) foi aplicado.

Tomaremos como ponto de partida a densidade Lagrangeana

\begin{equation}
\mathcal{L}=-\,\frac{1}{4}F_{\mu \,\nu }\,F^{\mu \,\nu }+m\varepsilon
^{\alpha \,\beta \,\gamma }\,A_{\alpha }\,\partial _{\beta }\,A_{\gamma },
\label{327}
\end{equation}
\bigskip onde a m\'{e}trica $\left( -1,1,1\right) $ \'{e} adotada.$%
\smallskip \medskip $

A Hamiltoniana generalizada tem a seguinte forma can\^{o}nica:

\begin{equation}
\mathcal{H}=\int d^{\,2}\,x\,\left[ \frac{1}{2}\pi \,^{i}\,\pi \,^{i}+\frac{1%
}{2}\left( \varepsilon ^{i\,j}\,\partial ^{\,i}\,A^{\,j}\right) ^{2}+\frac{1%
}{2}m^{2}\,A^{\,k}\,A^{\,k}+m\varepsilon \,^{i\,j}\,A^{\,i}\,\pi ^{\,j}%
\right] ,  \label{328}
\end{equation}
com as rela\c{c}\~{o}es de v\'{\i}nculos de segunda-classe:
\begin{eqnarray}
\phi ^{1}&=&\pi ^{\,0}=0,  \label{329}\\
\phi ^{2}&=&\partial ^{\,i}\,\pi ^{\,i}+m\,\varepsilon ^{\,i\,j}\,\partial
^{\,j}\,A^{\,i}=0,  \label{330}\\
\phi ^{3}&=&A^{0}=0,  \label{331}\\
\phi ^{4}&=&\partial ^{\,i}\,A^{\,i}=0.  \label{332}
\end{eqnarray}
\medskip Para estabelecer uma estrutura simpl\'{e}tica, vamos renomear as
vari\'{a}veis de campo de acordo com a seguinte correspond\^{e}ncia:

\begin{equation}
\left( A^{0},\,A^{1},\,A^{2},\,\pi ^{0},\,\pi ^{1},\,\pi ^{2}\right)
\Leftrightarrow \left( \xi ^{1},\,\xi ^{2},\,\xi ^{3},\,\xi ^{4},\,\xi
^{5},\,\xi ^{6}\right) .  \label{333}
\end{equation}
\medskip Os v\'{\i}nculos \ $\phi ^{\,i}$ \ \ definem \ \ a \ m\'{e}trica
local, $\ \ J_{i\,j}$, \ \ que \'{e} a \ inversa de \ \ \ \ \ \ \
\smallskip\ $\ J^{i\,j}\,\left( x,y\right) =\left\{ \phi ^{\,i}\,\left(
x\right) ,\,\phi ^{\,j}\,\left( y\right) \,\right\} ,$ e formalmente se
escreve como abaixo:

\begin{equation}
J^{-1}=\left( 
\begin{array}{cccc}
0 & 0 & \delta ^{\,2}\,\left( x-y\right)  & 0 \\ 
0 & 0 & 0 & \nabla ^{-2} \\ 
-\,\delta ^{\,2}\,\left( x-y\right)  & 0 & 0 & 0 \\ 
0 & -\,\nabla ^{-2} & 0 & 0
\end{array}
\right) .  \label{334}
\end{equation}
\medskip Ap\'{o}s aplicar (\ref{12}), encontramos para a matriz do
projetor:\ \ 
\begin{equation}
\Lambda =\left( 
\begin{array}{cccccc}
0 & 0 & 0 & 0 & 0 & 0 \\ 
0 & \delta ^{2}\left( x-y\right) -\frac{\partial _{\,1}^{\,x}\,\partial
_{\,1}^{\,y}}{\nabla ^{\,2}} & -\,\frac{\partial _{1}^{x}\partial _{2}^{y}}{%
\nabla ^{2}} & 0 & 0 & 0 \\ 
0 & -\frac{\partial _{\,2}^{\,x}\,\partial _{\,1}^{\,y}}{\nabla ^{\,2}} & 
\delta ^{2}\left( x-y\right) -\frac{\partial _{\,2}^{\,x}\,\partial
_{\,2}^{\,y}}{\nabla ^{\,2}} & 0 & 0 & 0 \\ 
0 & 0 & 0 & 0 & 0 & 0 \\ 
0 & 0 & -\,m\,\delta ^{\,2}\left( x-y\right)  & 0 & \delta ^{2}\left(
x-y\right) -\frac{\partial \,_{1}^{x}\,\partial _{1}^{y}}{\nabla ^{\,2}} & -%
\frac{\partial \,_{1}^{x}\,\partial \,_{2}^{y}}{\nabla ^{2}} \\ 
0 & m\,\delta ^{\,2}\left( x-y\right)  & 0 & 0 & -\frac{\partial
\,_{2}^{x}\,\partial \,_{1}^{y}}{\nabla ^{2}} & \delta ^{2}\left( x-y\right)
-\frac{\partial \,_{2}^{x}\,\partial \,_{2}^{y}}{\nabla ^{2}}
\end{array}
\right)   \label{335}
\end{equation}
\ \ \ \ \ \ \ \ \ \ \ \ \ \ \ \ \ \ \ \ \ \ \ \ \ \ \ \ \ \ \ \ \ \ \ \ \ \
\ \ \ \ \ \ \ \ \bigskip \medskip\ \bigskip 

Obter as vari\'{a}veis f\'{\i}sicas,\thinspace\ $\xi _{\mu }^{\,\ast
}\,\left( x\right) $, \'{e} uma quest\~{a}o de aplicar a prescri\c{c}\~{a}o (%
\ref{13}); obtemos assim:\ \ 

\begin{equation}
\xi ^{\,1\ast }\,\left( x\right) =0,  \label{336.a}
\end{equation}

\begin{equation}
\xi ^{\,2\ast }\,\left( x\right) =A_{\,1}^{\,\perp }\,\left( x\right) ,
\label{336.b}
\end{equation}
\begin{equation}
\xi ^{\,3\ast }\left( x\right) =A_{\,2}^{\,\perp }\,\left( x\right) ,
\label{336.c}
\end{equation}

\begin{equation}
\xi ^{\,4\ast }\,\left( x\right) =0,  \label{336.d}
\end{equation}

\begin{equation}
\xi ^{\,5\ast }\,\left( x\right) =\pi _{\,1}^{\,\perp }\,\left( x\right)
-m\,A_{\,2}^{\,\perp }\,\left( x\right) ,  \label{336.e}
\end{equation}

\begin{equation}
\xi ^{\,6\ast }\,\left( x\right) =\pi _{\,2}^{\,\perp }\,\left( x\right)
+m\,A_{\,1}^{\,\perp }\,\left( x\right) .  \label{336.f}
\end{equation}
\medskip

Por um lado, nossa Hamiltoniana vinculada original, escrita em coordenadas
simpl\'{e}ticas, toma a forma
\begin{equation}
\mathcal{H}=\int d^{\,2}\,x\,\left[ \,\frac{1}{2}\,\left( \xi \,_{5}^{2}+\xi
\,_{6}^{2}\right) +\frac{1}{2}\,\left( \partial _{1}\,\xi _{3}-\partial
_{2}\,\xi _{2}\right) ^{\,2}+\frac{1}{2}\,m^{2}\,\left( \xi \,_{2}^{2}+\xi
\,_{3}^{2}\right) +m\left( \xi \,_{2}\,\xi \,_{6}-\xi \,_{3}\,\xi
\,_{5}\right) \right] ;  \label{337}
\end{equation}
por outro lado,a Hamiltoniana projetada fica::

\begin{equation}
\mathcal{H}^{\ast }=\int d^{2}x\left[ \frac{1}{2}\left( \xi _{5}^{\ast
2}+\xi _{6}^{\ast 2}\right) +\frac{1}{2}\left( \partial _{1}\xi _{3}^{\ast
}-\partial _{2}\xi _{2}^{\ast }\right) ^{2}+\frac{1}{2}m^{2}\left( \xi
_{2}^{\ast 2}+\xi _{3}^{\ast 2}\right) +m\left( \xi _{2}^{\ast }\xi
_{6}^{\ast }-\xi _{3}^{\ast }\xi _{5}^{\ast }\right) \right] .  \label{338}
\end{equation}
\medskip Voltando \`{a} nota\c{c}\~{a}o original do espa\c{c}o de fase, com
ajuda das equa\c{c}\~{o}es (\ref{336.a}-\ref{336.f}), finalmente concluimos
que a Hamiltoniana projetada toma a forma mais familiar abaixo:

\begin{equation}
\mathcal{H}^{\ast }=\int d^{2}\,x\,\left[ \frac{1}{2}\,\left( \pi
_{i}^{\perp }\,\pi _{i}^{\perp }+4\,m^{2}A\,_{i}^{\perp }\,A\,_{i}^{\perp
}\right) +\frac{1}{2}\left( \,\epsilon ^{\,i\,j}\,\partial
_{\,i}\,A\,_{j}^{\perp }\right) ^{2}+2\,m\,\left( A_{1}^{\perp }\,\pi
\,_{2}^{\perp }-A\,_{2}^{\perp }\,\pi _{1}^{\perp }\right) \right] .
\label{339}
\end{equation}
\smallskip Esta \'{e} a Hamiltoniana de Chern-Simons-Maxwell escrita em
termos das assim chamadas express\~{o}es transversas, que concorda com os
resultados encontrados em \cite{devecchi} , embora a partir de uma linha
diferente de argumentos.\smallskip 

Achamos interessante enfatizar, neste est\'{a}gio, qual foi nossa verdadeira
motiva\c{c}\~{a}o ao realizar este trabalho: aplicar e checar o MPS, uma vez
em que este \'{e} aplicado \ em um modelo massivo invariante de gauge em 3D.
A concord\^{a}ncia com os resultados em \cite{devecchi} confirmava a
plausibilidade do m\'{e}todo, que j\'{a} havia sido checado de forma
positiva para modelos em 4 e 2 dimens\~{o}es.

Gostar\'{\i}amos ainda de fazer uma importante observa\c{c}\~{a}o a respeito
deste resultado: a Hamiltoniana f\'{\i}sica \'{e} aquela dada por (\ref{338}%
), uma vez que as vari\'{a}veis f\'{\i}sicas, aquelas respeitando
par\^{e}nteses de Poisson can\^{o}nicos, s\~{a}o as \ $\xi ^{\ast }$'s, \ e
n\~{a}o as familiares vari\'{a}veis de campos transversos. A \'{u}nica
raz\~{a}o para escrever $H^{\ast }$ como em (\ref{339}) foi a de estabelecer
uma ponte entre nossa abordagem e a terminologia usual.

Voltando \`{a}s equa\c{c}\~{o}es de movimento e utilizando a Hamiltoniana
f\'{\i}sica no contexto das equa\c{c}\~{o}es de Hamilton-Jacob, encontramos:

\begin{eqnarray}
\ddot{\xi}_{2}^{\ast }&=&-\,2\,m^{2}\,\xi \,_{2}^{\ast }+\,\partial
_{2}\,\partial _{2}\,\xi \,_{2}^{\ast }\,-\partial _{1\,}\partial _{2}\,\xi
\,_{3}^{\ast }-2\,m\,\xi \,_{6}^{\ast },  \label{340.a}
\\
\ddot{\xi}_{3}^{\ast }&=&-\,2\,m^{2}\,\xi \,_{3}^{\ast }+\partial
_{1}\,\partial _{1}\,\xi \,_{3}^{\ast }-\partial _{1}\,\partial _{2}\,\xi
\,_{2}^{\ast }-2\,m\,\xi \,_{5}^{\ast },  \label{340.b}
\\
\ddot{\xi}_{5}^{\ast }&=&\,-\,2\,m^{2}\,\xi \,_{5}^{\ast }+\partial
_{2\,}\partial _{2}\,\xi \,_{5}^{\ast }-\partial _{1}\,\partial _{2}\,\xi
\,_{6}^{\ast }+m\,\left[ \,2\,m^{2}-\nabla ^{2}\right] \,\xi _{3}^{\ast },
\label{340.c}
\\
\ddot{\xi}_{6}^{\ast }&=&-\,2\,m^{2}\,\xi _{6}^{\ast }+\partial _{1}\,\partial
_{1}\,\xi _{6}^{\ast }-\partial _{1}\,\partial _{2}\,\xi _{5}^{\ast }-m\,%
\left[ 2m^{2}-\nabla ^{2}\right] \,\xi _{2}^{\ast }.  \label{340.d}
\end{eqnarray}
Aparentemente, estas equa\c{c}\~{o}es podem parecer bastante estranhas; mas,
se voltamos \'{a} nota\c{c}\~{a}o familiar, por meio da correspond\^{e}ncia
entre os A's, $\pi $'s e $\xi $'s (\ref{336.a}-\ref{336.f}) podemos
coloc\'a-las na forma:\smallskip
\begin{eqnarray}
\left( \square +4\,m^{\,2}\right) \,A_{\,1}^{\,\perp }&=&-\,2\,m\,\pi
_{\,2}^{\,\perp },  \label{341.a}\\
\left( \square +4\,m^{\,2}\right) \,A_{\,2}^{\,\perp }&=&\,2\,m\,\pi
_{\,1}^{\,\perp },  \label{341.b}\\
\square \,\pi _{\,1}^{\,\perp }&=&0,  \label{341.c}\\
\square \,\pi _{\,2}^{\,\perp }&=&0,  \label{341.d}
\end{eqnarray}
que resultam em assegurar que
\begin{equation}
\square \,\left( \square +4\,m^{2}\right) \,A_{\,i\,}^{\,\perp }=0,\,\,\ \ \
\left( i=1,2\right) .  \label{342}
\end{equation}
Esta equa\c{c}\~{a}o garante que a excita\c{c}\~{a}o f\'{\i}sica \'{e} um
vetor transverso massivo $\left( p^{2}=4\,m\,^{2}\right) .$ O quantum sem
massa ($p^{2}$\smallskip $=0$) \'{e} esp\'{u}rio: n\~{a}o possui papel na din%
\^{a}mica e n\~{a}o corresponde a nenhum modo f\'{\i}sico. Realmente, ao
acoplar o propagador do campo $A_{\mu }$ a uma corrente externa conservada,
a amplitude corrente-corrente \'{e} tal que a parte imagin\'{a}ria de seu res%
\'{\i}duo, tomada no polo $p^{2}=0$, se anula, o que confirma que o \'{u}ltimo n%
\~{a}o corresponde a nenhuma excita\c{c}\~{a}o f\'{\i}sica. Por outro lado, o
polo n\~{a}o-trivial $p^{2}=4m^{2}$ produz um res\'{\i}duo positivo-definido,
que refor\c{c}a seu car\'{a}cter f\'{\i}sico como o \'{u}nico grau de
liberdade carregado pelo campo $A_{\mu }$.

Esta an\'{a}lise  pode ainda ser feita, de forma mais transparente, se
escolhermos, sem perda de generalidade, uma determinada dire\c{c}\~{a}o de
propaga\c{c}\~{a}o ( p. ex. , $\vec{k}=\left( k,0\right) $): neste caso, as
equa\c{c}\~{o}es (\ref{336.a}-\ref{336.f}) revelam explicitamente a exist%
\^{e}ncia de apenas duas vari\'{a}veis independentes ( uma vez que $\xi
^{\,2\ast }=0$ e $\xi ^{\,5\ast }=-m\xi ^{\,3\ast }$), como j\'{a} se
esperaria a partir da contagem de graus de liberdade, ou mesmo, do tra\c{c}o
da matriz do projetor. O \'{u}nico campo sobrevivente teria como equa\c{c}%
\~{a}o de movimento, simplesmente, 
\begin{equation}
\left( \square +4\,m^{\,2}\right) \,A_{\,2}^{\,\perp }=0.  \label{343}
\end{equation}

\section{\sc Modelo Abeliano de Chern-Simons Estendido}

Dando continuidade \`{a}s investiga\c{c}\~{o}es em teorias 3D, analisaremos
agora o modelo 4D de Cremmer-Scherk-Kalb-Ramond (\cite{kr},\cite{cs}) quando
submetido ao processo de redu\c{c}\~{a}o dimensional. Obt\'{e}m-se assim um
modelo extendido de gauge, Abeliano, com um termo de Chern-Simons acoplando
um par de potenciais de gauge (\cite{cchmn1},\cite{cchmn2}). A
exist\^{e}ncia de um campo com simetria de gauge n\~{a}o-usual nos leva,
como veremos, a um particular conjunto de gauge-fixing, revelando, por fim,
a aus\^{e}ncia de conte\'{u}do f\'{\i}sico para este campo\cite{alvaro}.

Nosso ponto de partida \'{e} a densidade Lagrangeana 
\begin{eqnarray}
\mathcal{L} &=&-\frac{1}{4}F^{\mu \nu }F_{\mu \nu }-\frac{1}{4}G^{\mu \nu
}G_{\mu \nu }+\frac{1}{2}\partial ^{\mu }\varphi \partial _{\mu }\varphi +%
\frac{1}{2}\partial ^{\mu }Z_{\mu }\partial ^{\nu }Z_{\nu }-m\left( \partial
^{\mu }Z_{\mu }\right) \varphi   \nonumber \\
&&+m\varepsilon ^{\mu \nu \rho }B_{\mu }\partial _{\nu }A_{\rho },
\label{344}
\end{eqnarray}
com 
\begin{equation}
F^{\mu \nu }=\partial ^{\mu }A^{\nu }-\partial ^{\nu }A^{\mu },
\label{345.a}
\end{equation}
\begin{equation}
G^{\mu \nu }=\partial ^{\mu }B^{\nu }-\partial ^{\nu }B^{\mu },
\label{345.b}
\end{equation}
e a m\'{e}trica $\eta ^{\mu \nu }=(+,-,-)$. Calculando os momenta
canonicamente conjugados, temos 
\begin{equation}
\pi ^{\mu }\equiv \frac{\delta \mathcal{L}}{\delta \left( \partial
_{0}A_{\mu }\right) }=-F^{0\mu }+m\varepsilon ^{\nu 0\mu }B_{n}\;,
\label{346}
\end{equation}
que implicam em 
\begin{eqnarray}
\pi ^{0}&=&0  \label{347.a}\\
\pi ^{i}&=&-F^{0i}+m\varepsilon ^{0ik}B_{k}.  \label{347.b}
\end{eqnarray}
Tamb\'{e}m, 
\begin{equation}
P^{\mu }\equiv \frac{\delta \mathcal{L}}{\delta \left( \partial _{0}B_{\mu
}\right) }=-G^{0\mu },  \label{348}
\end{equation}
ou 
\begin{eqnarray}
P^{0}&=&0 \, ,  \label{349.a}\\
P^{i}&=&-G^{0i}.  \label{349.b}
\end{eqnarray}
\ \ \ \ Para o campo escalar, 
\begin{equation}
\pi _{\varphi }\equiv \frac{\delta \mathcal{L}}{\delta \left( \partial
_{0}\varphi \right) }=\partial _{0}\varphi .  \label{350}
\end{equation}
Finalmente, 
\begin{equation}
\pi ^{\prime \mu }\equiv \frac{\delta \mathcal{L}}{\delta \left( \partial
_{0}Z_{m}\right) }=\left( -m\varphi +\partial ^{\beta }Z_{\beta }\right)
\eta ^{\mu 0}  \label{351}
\end{equation}
ou 
\begin{equation}
\pi ^{\prime 0}=\left( +\partial ^{\beta }Z_{\beta }-m\varphi \right) 
\label{352.a}
\end{equation}
\begin{equation}
\pi ^{\prime i}=0.  \label{352.b}
\end{equation}

Podemos ent\~{a}o escrever a Hamiltoniana can\^{o}nica da teoria: 
\begin{eqnarray}
\mathcal{H}_{c} &=&\pi ^{\mu }\partial _{0}A_{\mu }+P^{\mu }\partial
_{0}B_{\mu }+\pi ^{\prime \mu }\partial _{0}Z_{\mu }+\pi _{\varphi }\partial
_{0}\varphi -\mathcal{L}  \nonumber \\
&=&\frac{1}{2}\pi _{i}^{2}+\frac{1}{2}P_{i}^{2}+\frac{1}{2}\pi _{\varphi
}^{2}+\frac{1}{2}\pi _{0}^{\prime 2}+A_{0}\left( \partial _{i}\pi
_{i}\right) +B_{0}\left( \partial _{i}P_{i}-m\varepsilon _{0ij}\partial
_{i}A_{j}\right) +  \nonumber \\
&&+\frac{1}{4}F_{ij}F_{ij}+m\varepsilon _{0ik}\pi _{i}B_{k}+\frac{1}{2}%
m^{2}B_{k}B_{k}+\frac{1}{4}G_{ij}G_{ij}+\frac{1}{2}\left( \partial
_{i}\varphi \right) ^{2}+\frac{1}{2}m^{2}\varphi ^{2}  \nonumber \\
&&+\pi _{0}^{\prime }\left( m\varphi +\partial _{i}Z_{i}\right) \;.
\label{353}
\end{eqnarray}
A Hamiltoniana prim\'{a}ria \'{e} ent\~{a}o 
\begin{equation}
\mathcal{H}_{p}=\mathcal{H}_{c}+v_{i}\phi _{i},  \label{354}
\end{equation}
onde os v\'{\i}nculos prim\'{a}rios s\~{a}o 
\begin{eqnarray}
\phi _{1} &=&\pi _{0}\approx 0  \nonumber \\
\phi _{2} &=&P_{0}\approx 0  \nonumber \\
\phi _{3} &=&\pi _{1}^{\prime }\approx 0  \nonumber \\
\phi _{4} &=&\pi _{2}^{\prime }\approx 0.  \label{355}
\end{eqnarray}
Como se pode notar imediatamente, $\phi _{3}$ e $\phi _{4}$ s\~{a}o a
primeira evid\^{e}ncia do caracter n\~{a}o usual dos $Z_{\mu }$'s a ser
posto \`{a} tona no conjunto de v\'{\i}nculos.As condi\c{c}\~{o}es de
consist\^{e}ncia impostas sobre este conjunto fornecem os v\'{\i}nculos
secund\'{a}rios: 
\begin{eqnarray}
\phi _{5} &=&\partial _{i}\pi _{i}\approx 0  \nonumber \\
\phi _{6} &=&\partial _{i}P_{i}-m\varepsilon _{0ij}\partial _{i}A_{j}\approx
0  \nonumber \\
\phi _{7} &=&\pi _{0}^{\prime }-f\left( t\right) \approx 0,  \label{356}
\end{eqnarray}
para uma fun\c{c}\~{a}o arbitr\'{a}ria $f(t)$. Estes s\~{a}o todos
v\'{\i}nculos de primeira-classe; as condi\c{c}\~{o}es de gauge-fixing
s\~{a}o escolhidas de modo que 
\begin{eqnarray}
\phi _{8} &=&A_{0}\approx 0  \nonumber \\
\phi _{9} &=&B_{0}\approx 0  \nonumber \\
\phi _{10} &=&Z_{1}\approx 0  \nonumber \\
\phi _{11} &=&Z_{2}\approx 0  \nonumber \\
\phi _{12} &=&\partial _{i}A_{i}\approx 0  \nonumber \\
\phi _{13} &=&\partial _{i}B_{i}\approx 0  \nonumber \\
\phi _{14} &=&Z_{0}\approx 0,  \label{357}
\end{eqnarray}
onde impomos o gauge-fixing de ``Coulomb'' sobre $Z_{\mu }$ em estreita
analogia com o procedimento usual como aplicado sobre $A_{\mu }$ (e $B_{\mu
} $). Como resultado l\'{\i}quido, a cole\c{c}\~{a}o de v\'{\i}nculos
relacionada a $Z_{\mu }$ j\'{a} indicam que suas vari\'{a}veis de espa\c{c}o
de fase devam ser exclu\'{\i}das do subconjunto din\^{a}mico.\bigskip

Vamos agora construir a matriz com elementos $g_{ij}(x,y)=\left\{ \Omega _{i}(x),\Omega
_{j}(y)\right\} $. Adotando a nota\c{c}\~{a}o convencional $\delta \equiv
\delta ^{2}\left( x-y\right) $, temos que $g(x,y)$ \'e dada por
\begin{eqnarray}
%&& g(x,y) = \nonumber \\ &&
\left( 
\begin{array}{cccccccccccccc}
0 & 0 & 0 & 0 & 0 & 0 & 0 & -\delta & 0 & 0 & 0 & 0 & 0 & 0 \\ 
0 & 0 & 0 & 0 & 0 & 0 & 0 & 0 & -\delta & 0 & 0 & 0 & 0 & 0 \\ 
0 & 0 & 0 & 0 & 0 & 0 & 0 & 0 & 0 & \delta & 0 & 0 & 0 & 0 \\ 
0 & 0 & 0 & 0 & 0 & 0 & 0 & 0 & 0 & 0 & \delta & 0 & 0 & 0 \\ 
0 & 0 & 0 & 0 & 0 & 0 & 0 & 0 & 0 & 0 & 0 & \partial _{i}^{x}\partial
_{i}^{y}\delta & 0 & 0 \\ 
0 & 0 & 0 & 0 & 0 & 0 & 0 & 0 & 0 & 0 & 0 & 0 & \partial _{i}^{x}\partial
_{i}^{y}\delta & 0 \\ 
0 & 0 & 0 & 0 & 0 & 0 & 0 & 0 & 0 & 0 & 0 & 0 & 0 & -\delta \\ 
\hspace{0.05in}\delta \hspace{0.05in} & 0 & 0 & 0 & 0 & 0 & 0 & 0 & 0 & 0 & 0
& 0 & 0 & 0 \\ 
0 & \hspace{0.05in}\delta \hspace{0.05in} & 0 & 0 & 0 & 0 & 0 & 0 & 0 & 0 & 0
& 0 & 0 & 0 \\ 
0 & 0 & \hspace{0.05in}-\delta \hspace{0.05in} & 0 & 0 & 0 & 0 & 0 & 0 & 0 & 
0 & 0 & 0 & 0 \\ 
0 & 0 & 0 & \hspace{0.05in}-\delta \hspace{0.05in} & 0 & 0 & 0 & 0 & 0 & 0 & 
0 & 0 & 0 & 0 \\ 
0 & 0 & 0 & 0 & -\partial _{i}^{x}\partial _{i}^{y}\delta & 0 & 0 & 0 & 0 & 0
& 0 & 0 & 0 & 0 \\ 
0 & 0 & 0 & 0 & 0 & -\partial _{i}^{x}\partial _{i}^{y}\delta & 0 & 0 & 0 & 0
& 0 & 0 & 0 & 0 \\ 
0 & 0 & 0 & 0 & 0 & 0 & \hspace{0.05in}\delta \hspace{0.05in} & 0 & 0 & 0 & 0
& 0 & 0 & 0
\end{array}
\right) \nonumber  \label{358} \\
&& 
\end{eqnarray}
cuja inversa, $g^{-1}(x,y)$,  \'e dada por
\begin{eqnarray}
%&& =  \nonumber \\ &&
\left( 
\begin{array}{cccccccccccccc}
0 & 0 & 0 & 0 & 0 & 0 & 0 & \hspace{0.05in}\delta \hspace{0.05in} & 0 & 0 & 0
& 0 & 0 & 0 \\ 
0 & 0 & 0 & 0 & 0 & 0 & 0 & 0 & \hspace{0.05in}\delta \hspace{0.05in} & 0 & 0
& 0 & 0 & 0 \\ 
0 & 0 & 0 & 0 & 0 & 0 & 0 & 0 & 0 & \hspace{0.05in}-\delta \hspace{0.05in} & 
0 & 0 & 0 & 0 \\ 
0 & 0 & 0 & 0 & 0 & 0 & 0 & 0 & 0 & 0 & \hspace{0.05in}-\delta \hspace{0.05in%
} & 0 & 0 & 0 \\ 
0 & 0 & 0 & 0 & 0 & 0 & 0 & 0 & 0 & 0 & 0 & +\nabla ^{-2} & 0 & 0 \\ 
0 & 0 & 0 & 0 & 0 & 0 & 0 & 0 & 0 & 0 & 0 & 0 & +\nabla ^{-2} & 0 \\ 
0 & 0 & 0 & 0 & 0 & 0 & 0 & 0 & 0 & 0 & 0 & 0 & 0 & \hspace{0.05in}\delta 
\hspace{0.05in} \\ 
-\delta & 0 & 0 & 0 & 0 & 0 & 0 & 0 & 0 & 0 & 0 & 0 & 0 & 0 \\ 
0 & -\delta & 0 & 0 & 0 & 0 & 0 & 0 & 0 & 0 & 0 & 0 & 0 & 0 \\ 
0 & 0 & \delta & 0 & 0 & 0 & 0 & 0 & 0 & 0 & 0 & 0 & 0 & 0 \\ 
0 & 0 & 0 & \delta & 0 & 0 & 0 & 0 & 0 & 0 & 0 & 0 & 0 & 0 \\ 
0 & 0 & 0 & 0 & -\nabla ^{-2} & 0 & 0 & 0 & 0 & 0 & 0 & 0 & 0 & 0 \\ 
0 & 0 & 0 & 0 & 0 & -\nabla ^{-2} & 0 & 0 & 0 & 0 & 0 & 0 & 0 & 0 \\ 
0 & 0 & 0 & 0 & 0 & 0 & -\delta & 0 & 0 & 0 & 0 & 0 & 0 & 0
\end{array}
\right)  \label{359} \nonumber \\
&&
\end{eqnarray}

Vamos etiquetar os campos e seus correspondentes momenta como segue:
\begin{eqnarray}
\left( A^{0},A^{1},A^{2},\varphi ,B^{0},B^{1},B^{2},Z^{0},Z^{1},Z^{2},\pi
_{0},\pi _{1},\pi _{2},\pi _{\varphi },P_{0},P_{1},P_{2},\pi _{0}^{\prime
},\pi _{1}^{\prime },\pi _{2}^{\prime }\right) &\equiv &  \nonumber \\
\left( \xi _{1},\xi _{2},\xi _{3},\xi _{4},\xi _{5},\xi _{6},\xi _{7},\xi
_{8},\xi _{9},\xi _{10},\xi _{11},\xi _{12},\xi _{13},\xi _{14},\xi
_{15},\xi _{16},\xi _{17},\xi _{18},\xi _{19},\xi _{20}\right) . && 
\nonumber
\end{eqnarray}
Temos assim todos os ingredientes para calcular a matriz do projetor
simpl\'{e}tico (\ref{12}); a fim de evitar o trato de express\~{o}es
extensas e cansativas... vamos expressar diretamente as coordenadas
projetadas obtidas via (\ref{13}) aquelas n\~{a}o-nulas s\~{a}o: 
\begin{eqnarray}
\xi ^{2\ast }\left( x\right) &=& A^{1\perp }\left( x\right)  \label{360.a}
\\
\xi ^{3\ast }\left( x\right) &=& A^{2\perp }\left( x\right)  \label{360.b}
\\
\xi ^{4\ast }\left( x\right) &=& \varphi \left( x\right)  \label{360.c}
\\
\xi ^{6\ast }\left( x\right) &=& B^{1\perp }\left( x\right)  \label{360.d}
\\
\xi ^{7\ast }\left( x\right) &=& B^{2\perp }\left( x\right)  \label{360.e}
\end{eqnarray}
\begin{eqnarray}
\xi ^{12\ast }\left( x\right) &=&\pi _{1}^{\perp }\left( x\right) +m\int
d^{2}y\partial _{x_{2}}\nabla ^{-2}\left( x,y\right) \left[ \partial
_{y_{1}}B_{1}\left( y\right) +\partial _{y_{2}}B_{2}\left( y\right) \right] 
\nonumber \\
&=&\overline{\pi }_{1}^{\perp }\left( x\right)  \label{360.f}
\end{eqnarray}
\begin{eqnarray}
\xi ^{13\ast }\left( x\right) &=&\pi _{2}^{\perp }\left( x\right) -m\int
d^{2}y\partial _{x_{1}}\nabla ^{-2}\left( x,y\right) \left[ \partial
_{y_{1}}B_{1}\left( y\right) +\partial _{y_{2}}B_{2}\left( y\right) \right] 
\nonumber \\
&=&\overline{\pi }_{2}^{\perp }\left( x\right)  \label{360.g}
\end{eqnarray}
\begin{equation}
\xi ^{14\ast }\left( x\right) =\pi _{\varphi }\left( x\right)  \label{360.h}
\end{equation}
\begin{eqnarray}
\xi ^{16\ast }\left( x\right) &=&P_{1}^{\perp }\left( x\right) +m\partial
_{x_{1}}\int d^{2}y\nabla ^{-2}\left( x,y\right) \left[ \partial
_{y_{1}}A_{2}\left( y\right) -\partial _{y_{2}}A_{1}\left( y\right) \right] 
\nonumber \\
&=&P_{1}^{\perp }\left( x\right) +m\partial _{x_{1}}\int d^{2}y\nabla
^{-2}\left( x,y\right) \left[ \mathbf{\nabla \times }A^{\perp }\right] _{y}
\label{360.i}
\end{eqnarray}
\begin{eqnarray}
\xi ^{17\ast }\left( x\right) &=&P_{2}^{\perp }\left( x\right) +m\partial
_{x2}\int d^{2}y\nabla ^{-2}\left( x,y\right) \left[ \partial
_{y_{1}}A_{2}\left( y\right) -\partial _{y_{2}}A_{1}\left( y\right) \right] 
\nonumber \\
&=&P_{2}^{\perp }\left( x\right) +m\partial _{x_{2}}\int d^{2}y\nabla
^{-2}\left( x,y\right) \left[ \mathbf{\nabla \times }A^{\perp }\right] _{y}
\label{360.j}
\end{eqnarray}

A Hamiltoniana can\^{o}nica reduzida \'{e} obtida da Hamiltoniana prim\'{a}%
ria tomando em conta os v\'{\i}nculos e condi\c{c}\~{o}es de gauge-fixing,
vistos agora como igualdades fortes. Obtemos: 
\begin{eqnarray}
\mathcal{H}_{c}^{r} &=&\frac{1}{2}\pi _{i}^{2}+\frac{1}{2}P_{i}^{2}+\frac{1}{%
2}\pi _{\varphi }^{2}+\frac{1}{2}\pi _{0}^{\prime 2}+\frac{1}{4}%
F_{ij}F_{ij}+m\varepsilon _{0ik}\pi _{i}B_{k}+\frac{1}{2}m^{2}B_{k}B_{k}+ 
\nonumber \\
&&+\frac{1}{4}G_{ij}G_{ij}+\frac{1}{2}\left( \partial _{i}\varphi \right)
^{2}+\frac{1}{2}m^{2}\varphi ^{2}+\pi _{0}^{\prime }\left( m\varphi \right) ,
\label{361}
\end{eqnarray}
ou, em nota\c{c}\~{a}o simpl\'{e}tica, 
\begin{eqnarray}
\mathcal{H}_{c}^{r} &=&\frac{1}{2}\left( \xi _{12}^{2}+\xi _{13}^{2}\right) +%
\frac{1}{2}\left( \xi _{16}^{2}+\xi _{17}^{2}\right) +\frac{1}{2}\xi
_{14}^{2}+\frac{1}{2}\xi _{18}^{2}+\frac{1}{2}\left( \partial _{1}\xi
_{3}\right) ^{2}+\frac{1}{2}\left( \partial _{2}\xi _{2}\right) ^{2}+ 
\nonumber \\
&&-\left( \partial _{1}\xi _{3}\right) \left( \partial _{2}\xi _{2}\right)
-m\left( \xi _{12}\xi _{7}-\xi _{13}\xi _{6}\right) +\frac{1}{2}m^{2}\left(
\xi _{6}^{2}+\xi _{7}^{2}\right) +\frac{1}{2}\left( \partial _{1}\xi
_{7}\right) ^{2}+\frac{1}{2}\left( \partial _{2}\xi _{6}\right) ^{2}+ 
\nonumber \\
&&-\left( \partial _{1}\xi _{7}\right) \left( \partial _{2}\xi _{6}\right) +%
\frac{1}{2}\left( \partial _{1}\xi _{4}\right) ^{2}+\frac{1}{2}\left(
\partial _{2}\xi _{4}\right) ^{2}+\frac{1}{2}m^{2}\xi _{4}^{2}+\xi
_{18}\left( m\xi _{4}\right) .  \label{362}
\end{eqnarray}
A densidade Hamiltoniana f\'{\i}sica \'{e} obtida reescrevendo a express\~{a}%
o acima em termos das vari\'{a}veis projetadas: 
\begin{eqnarray}
\mathcal{H}^{\ast } &=&\frac{1}{2}\left( \xi _{12}^{\ast 2}+\xi _{13}^{\ast
2}\right) +\frac{1}{2}\left( \xi _{16}^{\ast 2}+\xi _{17}^{\ast 2}\right) +%
\frac{1}{2}\xi _{14}^{\ast 2}+\frac{1}{2}\left( \partial _{1}\xi _{3}^{\ast
}\right) ^{2}+\frac{1}{2}\left( \partial _{2}\xi _{2}^{\ast }\right) ^{2}+ 
\nonumber \\
&&-\left( \partial _{1}\xi _{3}^{\ast }\right) \left( \partial _{2}\xi
_{2}^{\ast }\right) -m\left( \xi _{12}^{\ast }\xi _{7}^{\ast }-\xi
_{13}^{\ast }\xi _{6}^{\ast }\right) +\frac{1}{2}m^{2}\left( \xi _{6}^{\ast
2}+\xi _{7}^{\ast 2}\right) +\frac{1}{2}\left( \partial _{1}\xi _{7}^{\ast
}\right) ^{2}+  \nonumber \\
&&+\frac{1}{2}\left( \partial _{2}\xi _{6}^{\ast }\right) ^{2}-\left(
\partial _{1}\xi _{7}^{\ast }\right) \left( \partial _{2}\xi _{6}^{\ast
}\right) +\frac{1}{2}\left( \partial _{1}\xi _{4}^{\ast }\right) ^{2}+\frac{1%
}{2}\left( \partial _{2}\xi _{4}^{\ast }\right) ^{2}+\frac{1}{2}m^{2}\xi
_{4}^{\ast 2}\;.  \label{363}
\end{eqnarray}

Finalmente, as equa\c{c}\~{o}es de movimento s\~{a}o obtidads diretamente
das equa\c{c}\~{o}es de Hamilton-Jacobi atrav\'{e}s dos par\^{e}nteses de
Poisson entre as vari\'{a}veis projetadas e a Hamiltoniana $\int d^{2}y\;%
\mathcal{H}^{\ast }(y)$. Assim procedendo obtemos: 
\begin{equation}
{\stackrel{.}{\xi }}_{4}^{\ast }(x)=\int d^{2}y\left\{ \xi _{4}^{\ast }(x),%
\mathcal{H}^{\ast }(y)\right\} =\xi _{14}^{\ast }(x)\;;  \label{364}
\end{equation}
esta fornece 
\begin{equation}
{\stackrel{..}{\xi }}_{4}^{\ast }={\stackrel{.}{\xi }}_{14}^{\ast }=\int
d^{2}y\left\{ \xi _{14}^{\ast },\mathcal{H}^{\ast }(y)\right\} =-m^{2}\xi
_{4}^{\ast }+\nabla ^{2}\xi _{4}^{\ast }\;,  \label{365}
\end{equation}
ou 
\begin{equation}
(\square +m^{2})\xi _{4}^{\ast }=0\;.  \label{366}
\end{equation}
Analogamente, obtemos 
\begin{eqnarray}
{\stackrel{..}{\xi }}_{2}^{\ast } &=&\partial _{2}\partial _{2}\xi
_{2}^{\ast }-\partial _{1}\partial _{2}\xi _{3}^{\ast }-m\xi _{17}^{\ast } 
\nonumber \\
{\stackrel{..}{\xi }}_{3}^{\ast } &=&\partial _{1}\partial _{1}\xi
_{3}^{\ast }-\partial _{1}\partial _{2}\xi _{2}^{\ast }+m\xi _{16}^{\ast } 
\nonumber \\
{\stackrel{..}{\xi }}_{6}^{\ast } &=&-m^{2}\xi _{6}^{\ast }+\partial
_{2}\partial _{2}\xi _{6}^{\ast }-\partial _{1}\partial _{2}\xi _{7}^{\ast
}-m\xi _{13}^{\ast }  \nonumber \\
{\stackrel{..}{\xi }}_{7}^{\ast } &=&-m^{2}\xi _{7}^{\ast }+\partial
_{1}\partial _{1}\xi _{7}^{\ast }-\partial _{1}\partial _{2}\xi _{6}^{\ast
}+m\xi _{12}^{\ast }.  \label{367}
\end{eqnarray}
Agora, as equa\c{c}\~{o}es (\ref{367}) podem ser reexpressas numa forma bem
mais simples se es\-co\-lhe\-mos, sem perda de generalidade, o momentum apontando
segundo o eixo x ($\vec{k}=(k,0)$), selecionando as componentes ($%
A_{2},B_{2} $) como aquelas transversas. Pode-se notar facilmente que tal
escolha acarreta no cancelamento das vari\'{a}veis $\xi _{2}^{\ast }$, $\xi
_{6}^{\ast }$ e $\xi _{12}^{\ast }$, e converte as vari\'{a}veis $\xi
_{16}^{\ast }$ e $\xi _{17}^{\ast }$ em: 
\begin{eqnarray}
\xi _{16}^{\ast } &=&-m\xi _{3}^{\ast },  \nonumber \\
\xi _{17}^{\ast } &=&P_{2}^{\perp }.  \nonumber
\end{eqnarray}

O conjunto de vari\'{a}veis independentes seria especificado,
correspondentemente, pelos pares ($\xi _{3}^{\ast },\xi _{13}^{\ast }$), ($%
\xi _{4}^{\ast },\xi _{14}^{\ast }$) e ($\xi _{7}^{\ast },\xi _{17}^{\ast }$%
). As equa\c{c}\~{o}es de movimento resultantes seriam ent\~{a}o: 
\begin{eqnarray}
\square \xi _{4}^{\ast } &=&-m^{2}\xi _{4}^{\ast },  \nonumber \\
\square \xi _{3}^{\ast } &=&-m^{2}\xi _{3}^{\ast },  \nonumber \\
\square \xi _{7}^{\ast } &=&-m^{2}\xi _{7}^{\ast }.  \label{368}
\end{eqnarray}

Temos ent\~{a}o a presen\c{c}a de um campo escalar massivo, $\xi _{4}^{\ast }
$,e dois campos vetorias transversos massivos, $\xi _{3}^{\ast }$ e $\xi
_{7}^{\ast }$, de acordo com o que seria esperado da contagem de graus de
liberdade tendo em conta a Lagrangeana 3D (\ref{344}) e os v\'{\i}nculos de
segunda classe presentes. Tal resultado \'{e} tamb\'{e}m compativel com a
aloca\c{c}\~{a}o natural de graus de liberdade f\'{\i}sicos que podem ser
inferidos do modelo original 4D. Al\'{e}m disso, o fato de que os dois
vetores introduzem no setor f\'{\i}sico contribui\c{c}\~{o}es transversas
massivas equivalentes sugere a possibilidade de se realizar o mapeamento consistente em um
novo modelo, procedimento este que pode ser implementado atrav\'{e}s da
identifica\c{c}\~{a}o de ambos os campos vetoriais (e parceiros, num contexto
supersim\'{e}trico), como proposto em \cite{cchmn1}.

\chapter{{\sc Cordas Bos\^{o}nicas na Presen\c{c}a de um Campo de
Kalb-Ramond}}

Vamos por fim analisar uma teoria em espa\c{c}o-tempo d-dimensional,
considerando um modelo de cordas bos\^{o}nicas abertas com extremidades
ligadas a D-branas sujeitas a um potencial de gauge $A^{i}\left( X\right) $
com field-strength constante $F_{ij}$, m\'{e}trica constante $g_{ij}$ e um
campo de Kalb-Ramond constante $B_{ij}\left( X\right) .$ A an\'{a}lise
Hamiltoniana de tais modelos tem sido feita em diversas variantes da
abordagem de Dirac (\cite{witten}-\cite{rudy}).Vamos construir o projetor
simpl\'{e}tico atrav\'{e}s dos elementos da matriz de Dirac (\ref{m11}),
encontrar as coordenadas n\~{a}o-vinculadas das cordas abertas e a
Hamiltoniana que governa sua din\^{a}mica. Veremos que se escolhermos as
vari\'{a}veis dependentes como aquelas que correspondem \`{a}s duas
extremidades das cordas, ent\~{a}o a corda aberta \'{e} equivalente a um
sistema com coordenadas c\'{\i}clicas\cite{mmi}.

A a\c{c}\~{a}o do sistema \'{e} dada pela seguinte rela\c{c}\~{a}o: 
\begin{eqnarray}
S &=&\frac{1}{4\pi \alpha ^{\prime }}\int d^{2}\sigma \left[ \partial
_{a}X^{i}\partial ^{a}X_{i}+2\pi \alpha ^{\prime }B_{ij}\varepsilon
^{ab}\partial _{a}X^{i}\partial _{b}X^{j}\right]   \nonumber \\
&&+\int d\sigma \delta (\sigma -\pi )A_{i}\partial _{\sigma }X^{i}-\int
d\sigma \delta (\sigma )A_{i}\partial _{\sigma }X^{i},  \label{369}
\end{eqnarray}
onde $\sigma ^{\alpha }=(\tau ,\sigma )$ s\~{a}o as coordenadas ``folha do
mundo'' e $\epsilon ^{ab}$ \'{e} o s\'{\i}mbolo anti-sim\'{e}trico em duas
dimens\~{o}es $a,b=0,1$. Escolhemos por conveni\^{e}ncia trabalhar em um espa%
\c{c}o-tempo Euclidiano $d$-dimensional e os ``target-space indices'' s\~{a}%
o $i,j=1,2,\ldots ,d$, onde $d$ \'{e} par e para corda cr\'{\i}tica $d=26$.
Uma vez que o campo de gauge \'{e} constante, pode-se convenientemente faz%
\^{e}-lo nulo $F_{ij}=0$. Variando a a\c{c}\~{a}o (\ref{369}) com respeito
aos campos obtemos as equa\c{c}\~{o}es de movimento 
\begin{equation}
\partial _{a}\partial ^{a}X^{i}=0,  \label{370}
\end{equation}
que valem se, e apenas se, as condi\c{c}\~{o}es de contorno mixtas de
Dirichlet e Neumann valem 
\begin{equation}
g_{ij}\partial _{\sigma }X^{j}+2\pi \alpha ^{\prime }B_{ij}\partial _{\sigma
}X^{j}|_{\sigma =0,\pi }=0.  \label{371}
\end{equation}
As condi\c{c}\~{o}es de contorno (\ref{371}) representam rela\c{c}\~{o}es
entre os momenta da corda. No processo de quantiza\c{c}\~{a}o usual, estas
condi\c{c}\~{o}es devem ser impostas no espa\c{c}o de Hilbert da teoria.
Entretanto, existe uma abordagem alternativa para quantizar o sistema, na
qual as condi\c{c}\~{o}es de contorno s\~{a}o tratadas como v\'{\i}nculos alg%
\'{e}bricos (\cite{ardalan1}-\cite{chu1}).Para isto, deve-se primeiramente
discretizar a corda dividindo a extens\~{a}o do par\^{a}metro $\sigma $ por $%
\epsilon >0$. A coordenada $X^{i}(\sigma )$ para $i=1,2,\ldots ,d$ \'{e}
equivalente ao seguinte conjunto discreto $X_{\alpha }^{i}$ onde $\alpha
=1,2,\ldots ,m$ e $\epsilon =\pi /m$. \'{E} f\'{a}cil ver que a Lagrangeana
discretizada tem a seguinte forma 
\begin{equation}
L=\left( 4\pi \alpha ^{\prime }\right) ^{-1}\sum\limits_{\alpha }\left[
\varepsilon \left( \stackrel{.}{X}_{\alpha }^{i}\right) ^{2}-\frac{1}{%
\varepsilon }\left( X_{\alpha +1}^{i}-X_{\alpha }^{i}\right) ^{2}+4\pi
\alpha ^{\prime }B_{ij}\stackrel{.}{X}_{\alpha }^{i}\left( X_{\alpha
+1}^{j}-X_{\alpha }^{j}\right) \right] ,  \label{372}
\end{equation}
enquanto que as condi\c{c}\~{o}es de contorno (\ref{371}) em $\sigma =0$ se
tornam 
\begin{equation}
\frac{g_{ij}}{\varepsilon }\left( X_{2}^{j}-X_{1}^{j}\right) +2\pi \alpha
^{\prime }B_{ij}\stackrel{.}{X}_{1}^{j}=0.  \label{373}
\end{equation}
Similarmente, um conjunto id\^{e}ntico de condi\c{c}\~{o}es de contorno
discretizadas \'{e} obtido em $\sigma =\pi $ com $\{1,2\}$ em (\ref{371})
substituido por $\{m,m-1\}$.

Devido \`{a} discretiza\c{c}\~{a}o, as condi\c{c}\~{o}es de contorno s\~{a}o
equivalentes a v\'{\i}nculos alg\'{e}bricos sobre as extremidades da corda e
sobre seus primeiros vizinhos, enquanto que as vari\'{a}veis do centro s\~{a}%
o desvinculadas (\cite{ardalan1} e \cite{kim}). Se passamos \`{a}s vari\'{a}%
veis canonicamente conjugadas $X_{\alpha }^{i}$ e $P_{i\alpha }$, podemos
checar facilmente que os v'nculos, que toma a seguinte forma, 
\begin{equation}
w_{i}=\frac{1}{\varepsilon }\left[ \left( 2\pi \alpha ^{\prime }\right)
^{2}B_{ij}P_{1}^{j}-\left( 2\pi \alpha ^{\prime }\right)
^{2}B_{ij}B^{jk}\left( X_{2k}-X_{1k}\right) +g_{ij}\left(
X_{2}^{j}-X_{1}^{j}\right) \right] \approx 0,  \label{374}
\end{equation}
s\~{a}o de segunda classe. Existe um segundo conjunto de v\'{\i}nculos $\gamma
_{i}$ na outra extremidade da corda, os quais podem ser obtidos de (\ref{374}%
) apenas trocando as correspondentes coordenadas e momenta. Na an\'{a}lise
subsequente, podemos eliminar o par\^{a}metro $\epsilon $.

De forma a construir o projetor sobre a superf\'{\i}cie dos v\'{\i}nculos
vamos organizar as vari\'{a}veis matriciais $X_{\alpha }^{i}$ e $P_{i\alpha
} $ em um vetor coluna por $Z^{M}$. \'{E} f\'{a}cil perceber que a
cor\-res\-pon\-d\^{e}ncia entre os \'{\i}ndices $\{i,\alpha \}$ e $\{M\}$ \'{e}
dada pela seguinte rela\c{c}\~{a}o: 
\begin{equation}
M=\left( \mu +\alpha \right) +\left( m-1\right) \left( \mu -1\right) -1,
\label{375}
\end{equation}
onde $\mu =i,d+i$ e $i=1,2,\ldots ,d$. \'{E} \'{u}til tamb\'{e}m manter a
guarda sobre as coordenadas can\^{o}nicas e seus momenta conjugados nesta
nota\c{c}\~{a}o. De (\ref{375}) pode-se ver que a correspond\^{e}ncia entre
os dois conjuntos de vari\'{a}veis \'{e} 
\begin{equation}
X_{\alpha }^{i},~P_{\alpha i}\rightarrow Z^{m\left( i-1\right) +\alpha
},~Z_{m\left( d+i-1\right) +\alpha }  \label{376}
\end{equation}
Em seguida, vamos explicitar as vari\'{a}veis can\^{o}nicas em vari\'{a}veis
de extremos e vari\'{a}veis de centro, respectivamente, explicitando o
\'{\i}ndice $\alpha $ como segue: $\alpha =1,2,n,m-1,m$, onde $n=3,4,\ldots
,m-2$. \'{E} imediato computar os par\^{e}nteses de Dirac para o sistema
discretizado ( veja, por exemplo,\cite{kim}). A inversa da matriz dos
v\'{\i}nculos \'{e} dada pela seguinte rela\c{c}\~{a}o: 
\begin{equation}
\left( C^{-1}\right) ^{ij}=\frac{\varepsilon ^{2}}{2\left( 2\pi \alpha
^{\prime }\right) ^{2}}\left[ \frac{1}{\left( g+2\pi \alpha ^{\prime
}B\right) }\frac{1}{B}\frac{1}{\left( g-2\pi \alpha ^{\prime }B\right) }%
\right] ^{ij}.  \label{377}
\end{equation}

Com isto, podemos escrever a matriz $\Lambda $ usando a f\'{o}rmula (\ref
{m11}) com as vari\'{a}veis (\ref{376}). Para escrever seus elementos
n\~{a}o-nulos observamos de (\ref{375}) que o \'{\i}ndice $M$ pode ser
determinado por $i,\alpha $ e $d+i,\alpha $. \'{E} tamb\'{e}m importante
observar que a matriz do projetor consiste em quatro blocos de acordo com o
tipo de coordenadas $Z$'s sobre as quais ele atua e mapeia , i.e.,
coordenadas tipo-$X$ e tipo-$P$, respectivamente. As componentes
n\~{a}o-nulas do primeiro bloco da matriz $\Lambda $ s\~{a}o os seguintes: 
\begin{eqnarray}
2\Lambda _{\;j(1)}^{i(1)} &=&2\Lambda _{\;j(2)}^{i(1)}=\Lambda
_{\;j(2)}^{i(2)}=\Lambda _{\;j(m-1)}^{i(m-1)}=2\Lambda
_{\;j(m-1)}^{i(m)}=2\Lambda _{\;j(m)}^{i(m)}=\delta _{j}^{i}.  \nonumber \\
\Lambda _{\;j(n^{\prime })}^{i(n)} &=&\delta _{nn^{\prime }}\delta _{j}^{i}
\label{378}
\end{eqnarray}

Note que o \'{\i}ndice $\alpha $ n\~{a}o \'{e} um \'{\i}ndice covariante, e
portanto sua posi\c{c}\~{a}o relativa \'{e} irrelevante. Ele apenas marca as
diferentes coordenadas discretas de centro ou das extremidades da corda. O
bloco (\ref{378}) mapeia coordenadas de posi\c{c}\~{a}o em coordenadas de
posi\c{c}\~{a}o. Da mesma maneira podemos escrever os elementos n\~{a}%
o-nulos para os outros blocos: 
\begin{equation}
\Lambda ^{i(1)~d+j(1)}=-\Lambda ^{i(m)~d+j(m)}=\frac{1}{2}(2\pi \alpha
^{\prime })^{2}\left( \frac{1}{\left( g+2\pi \alpha ^{\prime }B\right) }B%
\frac{1}{\left( g-2\pi \alpha ^{\prime }B\right) }\right) ^{ij},  \label{379}
\end{equation}
para o bloco que mapeia $P$'s em $X$'s, 
\begin{eqnarray}
\Lambda _{d+i(1)~j(1)} &=&-\Lambda _{d+i(1)~j(2)}=\Lambda
_{d+i(2)~j(1)}=\Lambda _{d+i(2)~j(2)}=-\Lambda _{d+i(m-1)~j(m-1)}  \nonumber
\\
&=&-\Lambda _{d+i(m-1)~j(m)}=\Lambda _{d+i(m)~j(m-1)}=-\Lambda _{d+i(m)~j(m)}
\nonumber \\
&=&\frac{1}{2(2\pi \alpha ^{\prime })^{2}}\left( \left( g+2\pi \alpha
^{\prime }B\right) \frac{1}{B}\left( g-2\pi \alpha ^{\prime }B\right)
\right) _{ij},  \label{380}
\end{eqnarray}
para o bloco que mapeia $X$'s em $P$'s e 
\begin{eqnarray}
2\Lambda _{d+i(1)}^{\;\;\;d+j(1)} &=&2\Lambda
_{d+i(1)}^{\;\;\;d+j(2)}=\Lambda _{d+i(2)}^{\;\;\;d+j(2)}=\Lambda
_{d+i(m-1)}^{\;\;\;d+j(m-1)}=2\Lambda _{d+i(m)}^{\;\;\;d+j(m-1)}=2\Lambda
_{d+i(m)}^{\;\;\;d+j(m)}=\delta _{j}^{i},  \nonumber \\
\Lambda _{d+i(n)}^{\;\;\;d+j(n^{\prime })} &=&\delta _{nn^{\prime }}\delta
_{j}^{i}  \label{381}
\end{eqnarray}
para o bloco que mapeia $P$'s em $P$'s. Os elementos do projetor s\~{a}o
matrizes com elementos indexados por $i,j$. N\'{o}s escolhemos trabalhar com
a estrutura can\^{o}nica dos \'{\i}ndices, i.e. $(X^{i},P_{j})$. Ent\~{a}o 
\'{e} importante que a matriz $\Lambda $ mantenha esta estrutura atrav\'{e}s
da proje\c{c}\~{a}o. De outra forma ter\'{\i}amos que incluir nela a m\'{e}%
trica $g_{ij}$ para abaixar ou levantar \'{\i}ndices. Isto mudaria
ligeiramente a forma da matriz de Dirac. Estas complica\c{c}\~{o}es s\~{a}o
evitadas pela constru\c{c}\~{a}o acima e a matriz $\Lambda $ dada em (\ref
{378}-\ref{381}) \'{e} consistente com a covari\^{a}ncia do espa\c{c}o de
fase.

Ao atuar com a matriz $\Lambda $ sobre as coordenadas do espa\c{c}o de fase
chegamos \`{a}s vari\'{a}veis projetadas sobre a superf\'{\i}cie de
v\'{\i}nculos. De (\ref{378}-\ref{381}) obtemos as seguintes coordenadas
projetadas: 
\begin{eqnarray}
Z^{\star m\left( i-1\right) +1} &=&\frac{1}{2}Z^{m\left( i-1\right) +1}+%
\frac{1}{2}Z^{m\left( i-1\right) +2}  \nonumber \\
&&+\frac{1}{2}T^{-2}\left( C^{-1}BA^{-1}\right) ^{ij}Z_{m\left( d+j-1\right)
+1}  \nonumber \\
Z^{\star m\left( i-1\right) +2} &=&Z^{m\left( i-1\right) +2}  \nonumber \\
Z^{\star m\left( i-1\right) +n} &=&Z^{m\left( i-1\right) +n}  \nonumber \\
Z^{\star m\left( i-1\right) +\left( m-1\right) } &=&Z^{m\left( i-1\right)
+\left( m-1\right) }  \nonumber \\
Z^{\star m\left( i-1\right) +m} &=&\frac{1}{2}Z^{m\left( i-1\right) +m}+%
\frac{1}{2}Z^{m\left( i-1\right) +\left( m-1\right) }  \nonumber \\
&&-\frac{1}{2}T^{-2}\left( C^{-1}BA^{-1}\right) ^{ij}Z_{m\left( d+j-1\right)
+m}  \nonumber \\
Z_{m\left( d+i-1\right) +1}^{\star } &=&\frac{1}{2}Z_{m\left( d+i-1\right)
+1}-\frac{1}{2}T^{2}\left( CB^{-1}A\right) _{ij}Z^{m\left( j-1\right) +2} 
\nonumber \\
&&+\frac{1}{2}T^{2}\left( CB^{-1}A\right) _{ij}Z^{m\left( j-1\right) +1} 
\nonumber \\
Z_{m\left( d+i-1\right) +2}^{\star } &=&Z_{m\left( d+i-1\right) +2}+\frac{1}{%
2}T^{2}\left( CB^{-1}A\right) _{ij}Z^{m\left( j-1\right) +2}  \nonumber \\
&&+\frac{1}{2}T^{2}\left( CB^{-1}A\right) _{ij}Z^{m\left( j-1\right) +1}+%
\frac{1}{2}Z_{m\left( d+i-1\right) +1}  \nonumber \\
Z_{m\left( d+i-1\right) +n}^{\star } &=&Z_{m\left( d+i-1\right) +n} 
\nonumber \\
Z_{m\left( d+i-1\right) +m-1}^{\star } &=&Z_{m\left( d+i-1\right) +m-1}-%
\frac{1}{2}T^{2}\left( CB^{-1}A\right) _{ij}Z^{m\left( j-1\right) +m-1} 
\nonumber \\
&&+\frac{1}{2}T^{2}\left( CB^{-1}A\right) _{ij}Z^{m\left( j-1\right) +m}+%
\frac{1}{2}Z_{m\left( d+i-1\right) +m}  \nonumber \\
Z_{m\left( d+i-1\right) +m}^{\star } &=&-\frac{1}{2}Z_{m\left( d+i-1\right)
+m}-\frac{1}{2}T^{2}\left( CB^{-1}A\right) _{ij}Z^{m\left( j-1\right) +m} 
\nonumber \\
&&+\frac{1}{2}T^{2}\left( CB^{-1}A\right) _{ij}Z^{m\left( j-1\right) +m-1},
\label{382}
\end{eqnarray}

onde usamos as seguintes nota\c{c}\~{o}es abreviadas 
\begin{eqnarray}
A_{ij} &=&(g-2\pi \alpha ^{\prime }B)_{ij}  \nonumber \\
B_{ij} &=&B_{ij}  \nonumber \\
C_{ij} &=&(g+2\pi \alpha ^{\prime }B)_{ij}  \label{383}
\end{eqnarray}
e $T=(2\pi \alpha ^{\prime })^{-1}$. Note que nem todas as vari\'{a}veis em (%
\ref{382}) s\~{a}o independentes. Entretanto, as rela\c{c}\~{o}es entre elas
s\~{a}o lineares. Podemos ver que existem duas vari\'{a}veis dependentes
sobre a superf\'{\i}cie de v\'{\i}nculos. Usando a simetria com respeito \`{a}
troca das extremidades da corda encontra-se facilmente esta rela\c{c}\~{a}%
o. Podemos escolher as coordenadas nas extremidades como vari\'{a}veis
dependentes e obtemos as seguintes rela\c{c}\~{o}es para estas em termodas
coordenadas independentes: 
\begin{equation}
Z^{\star m\left( i-1\right) +1}=T^{-2}\left( C^{-1}BA^{-1}\right)
^{ij}Z_{m\left( d+i-1\right) +1}^{\star }+Z^{\star m\left( i-1\right) +2}
\label{384}
\end{equation}
\begin{equation}
Z^{\star m\left( i-1\right) +m}=-T^{-2}\left( C^{-1}BA^{-1}\right)
^{ij}Z_{m\left( d+i-1\right) +m}^{\star }+Z^{\star m\left( i-1\right) +m-1}.
\label{385}
\end{equation}

Das rela\c{c}\~{o}es acima, vemos que o sistema \'{e} equivalente a um
sistema com $2d(m-1)$ \ vari\'{a}veis independentes que parametrizam
localmente a superf\'{\i}cie de v\'{\i}nculos, o que corresponde ao
n\'{u}mero correto de graus de liberdade para $2dm$ vari\'{a}veis totais e $%
2d$ v\'{\i}nculos de segunda classe originais. Estas vari\'{a}veis comutam
entre s\'{\i} em concord\^{a}ncia com o resultado obtido em \cite{rudy}, mas
nesta descri\c{c}\~{a}o o sistema \'{e} c\'{\i}clico nas coordenadas das
extremidades, o que implica em que os correspondentes momenta s\~{a}o
conservados. A Hamiltoniana deste sistema tem a seguinte express\~{a}o 
\begin{eqnarray}
H^{\star } &=&\frac{1}{4\pi \alpha ^{\prime }\varepsilon }%
\sum\limits_{n=3}^{m-2}\left[ \left( 2\pi \alpha ^{\prime }\right)
^{2}\left( Z_{m\left( d+i-1\right) +n}^{\star }-B_{ij}\left( Z^{\star
m\left( j-1\right) +n+1}-Z^{\star m\left( j-1\right) +n}\right) \right)
^{2}\right. +  \nonumber \\
&&+\left. \left( Z^{\star m\left( j-1\right) +n+1}-Z^{\star m\left(
j-1\right) +n}\right) ^{2}\right] +  \nonumber \\
&&+\frac{4(\pi \alpha ^{\prime })^{3}}{\varepsilon }\left[ \left(
C^{-1}BA^{-1}\right) ^{ij}Z_{m\left( d+j-1\right) +m}^{\star }\right] ^{2}+%
\frac{4(\pi \alpha ^{\prime })^{3}}{\varepsilon }\left[ \left(
C^{-1}BA^{-1}\right) ^{ij}Z_{m\left( d+j-1\right) +1}^{\star }\right] ^{2}+ 
\nonumber \\
&&+\frac{\pi \alpha ^{\prime }}{\varepsilon }\left[ Z_{m\left( d+i-1\right)
+1}^{\star }+\frac{1}{\left( 2\pi \alpha ^{\prime }\right) ^{-2}}%
B_{ij}\left( C^{-1}BA^{-1}\right) ^{ij}Z_{m\left( d+j-1\right) +1}^{\star }%
\right] ^{2}+  \nonumber \\
&&+\frac{\pi \alpha ^{\prime }}{\varepsilon }\left[ Z_{m\left( d+i-1\right)
+m-1}^{\star }+B_{ij}\left( \left( 2\pi \alpha ^{\prime }\right) ^{2}\left(
C^{-1}BA^{-1}\right) ^{ij}Z_{m\left( d+j-1\right) +m}^{\star }\right) \right]
^{2}.  \label{386}
\end{eqnarray}

Observamos que as coordenadas dos extremos da corda n\~{a}o aparecem em (\ref
{386}). Entretanto, os correspondentes momenta aparecem nos \'{u}ltimos
quatro termos. Esta Hamiltoniana est\'{a} escrita em termos das
vari\'{a}veis independentes, e representa, portanto, o funcional natural
para a quantiza\c{c}\~{a}o.

\chapter{{\sc Considera\c c\~oes Finais e Perspectivas Futuras}}

\'{E} interessante notar, primeiramente, que todo o progresso obtido nos
\'{u}ltimos anos para o problema da formula\c{c}\~{a}o can\^{o}nica de
teorias sujeitas a v\'{\i}nculos tenha sido no sentido de aumentar o espa\c{c}%
o de fase original, mediante a introdu\c{c}\~{a}o de vari\'{a}veis extras e
de uma simetria mais poderosa (BRST), ao inv\'{e}s de, ao contr\'{a}rio,
reduzir o n\'{u}mero de vari\'{a}veis, eliminando aquelas esp\'{u}rias
atrav\'{e}s das fun\c{c}\~{o}es arbitr\'{a}rias introduzidas pelas simetrias
de gauge \cite{teitelboim}.

O m\'{e}todo dos projetores simpl\'{e}ticos tem como objetivo central a
identifica\c{c}\~{a}o do espa\c{c}o de fase reduzido, ou f\'{\i}sico. A
geometria simpl\'{e}tica determina que o conjunto de v\'{\i}nculos que
possibilita tal identifica\c{c}\~{a}o seja um conjunto m\'{\i}nimo de
v\'{\i}nculos independentes de segunda-classe. Tal restri\c{c}\~{a}o torna
necess\'{a}ria a solu\c{c}\~{a}o do problema do gauge-fixing naquelas
teorias em que existe a simetria de gauge. Sempre que tal quest\~{a}o pode
ser contornada ( como nos exemplos aqui tratados), a constru\c{c}\~{a}o do
projetor simpl\'{e}tico \'{e} formalmente assegurada. O vetor
simpl\'{e}tico, $\mathbf{\xi }$, uma vez projetado, tem algumas de suas
componentes identicamente nulas, enquanto que outras s\~{a}o linearmente
dependentes de um n\'{u}mero m\'{\i}nimo de coordenadas independentes, que
\'{e} o n\'{u}mero de graus de liberdade presentes. Esta identifica\c{c}%
\~{a}o das coordenadas linearmente independentes depende fortemente da
escolha das condi\c{c}\~{o}es de gauge-fixing. Como discutido ao final do
Cap\'{\i}tulo 2, em algumas situa\c{c}\~{o}es \'{e} poss\'{\i}vel que seja
necess\'{a}ria a investiga\c{c}\~{a}o dos auto-vetores da matriz $S$
associada ao projetor $\Lambda $, no lugar do projetor propriamente dito.
N\~{a}o encontramos, at\'{e} o momento, uma situa\c{c}\~{a}o em que tal fato
ocorresse, tendo sendo sempre poss\'{\i}vel identificar quais as coordenadas
independentes de forma trivial. Entretanto, consideramos que este deva ser
um ponto a ser melhor explorado em uma an\'{a}lise futura;talvez, mesmo sem
princ\'{\i}pios gerais ou considera\c{c}\~{o}es formais, fosse oportuno
encontrar exemplos de sistemas peculiares onde a simetria de gauge pedisse
que fossem explicitamente calculados os auto-vetores de que se falou acima.
Este pode ser o caso de teorias que descrevam spins mais altos em intera\c{c}%
\~{a}o, como nos modelos de supersimetria local estendida (N=2,4,8.).

A rela\c{c}\~{a}o (\ref{m11}), conectando o projetor simpl\'{e}tico \`{a}
matriz dos par\^{e}nteses de Dirac, tamb\'{e}m deve merecer, a nosso ver,
uma investiga\c{c}\~{a}o que forne\c{c}a mais clareza a respeito desta
conex\~{a}o.

Por fim, \ entre alguns problemas aos quais desejamos aplicar o m\'{e}todo
dos projetores, podemos citar uma extens\~{a}o daquele tratado no
Cap\'{\i}tulo 5, em que uma corda aberta com extremidades ligadas a D-branas
foi tratada \cite{mmi}: pretendemos considerar agora a situa\c{c}\~{a}o mais
interessante em que uma membrana aberta tem extremidades ligadas a outras
D-branas. Al\'{e}m de servir como mais um teste da aplicabilidade do
m\'{e}todo a objetos extensos, o tratamento de problemas como este pode
tamb\'{e}m contribuir para o esclarecimento de quest\~{o}es de interesse
bastante atuais. Felizmente, \'{e} bastante rico o universo de problemas em
que o m\'{e}todo presente pode ser utilizado com chances de trazer informa\c{%
c}\~{o}es \'{u}teis em suas investiga\c{c}\~{o}es.

\newpage

\section*{\sc Contribui\c c\~oes Cient\'{\i}ficas do Autor}

\begin{itemize}

\item ``On the Construction of U Matrix for QCD from Dirac Brackets'', 
J. C. de Mello, M. A. Santos and F. R. A. Sim\~ao, 
{\it J. Phys.} {\bf A21}, 493 (1987);

\item ``Physical Variables in Gauge Theories'', 
 J.~C.~de Mello, P.~Pitanga and M.~A.~Santos,
{\it Z. Phys.} {\bf C55} 271 (1992);

\item `` The Christ-Lee Model in the Approach of the Symplectic Projector Method'', 
J. C. de Mello, P. Pitanga and M. A. Santos,
{\it Braz. J. Phys.} {\bf 23} 214 (1993);

\item ``Sobre o Efeito Aharonov-Bohm'',
J. C. de Mello and M. A. Santos,
{\it Rev. Bras. Ens. F\'{\i}s.}, {\bf 4} 19 (1997);

\item ``Physical Variables for the Chern-Simons-Maxwell Theory without Matter'', 
J.~A.~He\-la\-y\"el-Neto and M.~A.~Santos, 
hep-th/9905065.

\item ``An Extended Abelian Chern-Simons Model and the Symplectic Projector Method'',
L.~R.~Manssur, A.~L.~Nogueira and M.~A.~Santos'', 
hep-th/0005214.

\item ``Unconstrained Variables of Non-Commutative Open Strings'', 
M.~A.~De Andrade, M.~A.~Santos and I.~V.~Vancea, {\it JHEP} {\bf 0106} 026 (2001), 
hep-th/0104154

\item ``Local Physical Coordinates from Symplectic Projector Method'', 
M.~A.~De Andrade, M.~A.~Santos and I.~V.~Vancea, 
{\it Mod. Phys. Lett.} {\bf A16} 1907 (2001), 
hep-th/0108197.

\item ``Sobre o Efeito Aharonov-Bohm'', 37$^a$ Reuni\~{a}o Anual da SBPC - Bras\'{\i}lia - (1985)

\item ``Obten\c{c}\~{a}o da Matriz U na QCD Atrav\'{e}s dos Par\^{e}nteses de Dirac'', 
VIII Encontro Nacional de F\'{\i}sica de Part\'{\i}culas e Campos - S\~ao Louren\c co - MG - (1987)

\item ``Probing the Symplectic Projector Method of Quantization in Planar Gauge Models'',
XXII Encontro Nacional de F\'{\i}sica de Part\'{\i}culas e Campos - S\~ao Louren\c co - MG - (2001) 

\item ``Free Variables on Constraint Surface for Non-Commutative Open Strings'',
XXII Encontro Nacional de F\'{\i}sica de Part\'{\i}culas e Campos - S\~ao Louren\c co - MG - (2001) 

\item ``M\'etodo dos Projetores Simpl\'eticos em Teorias de Campos de Gauge'',
XXII Encontro Nacional de F\'{\i}sica de Part\'{\i}culas e Campos - S\~ao Louren\c co - MG - (2001)

\end{itemize}


\begin{thebibliography}{99}

\bibitem{teitelboim}  M. Henneaux and C. Teitelboim, \textit{Quantization Of
Gauge Systems}, Princeton University Press, 1992\textit{\ }

\bibitem{cma}  C. Marcio do Amaral, \textit{Nuovo Cim. \textbf{B 25}}(1975)
817;

\bibitem{cma e pitanga}  C. Marcio do Amaral e P. Pitanga, \textit{Rev.
Bras. F\'{\i}sica }\textbf{3}(1982)473.

\bibitem{pitanga}  P. Pitanga, \textit{Nuovo Cim. }\textbf{A 103}(1990)1529.

\bibitem{dirac}  P.A.M. Dirac, \textit{Lectures on Quantum Mechanics},
Belfer Graduate School of Science, New York: Yeshiva University 1964.

\bibitem{sudarsham}  E.C.G. Sudarshan and N. Mukunda, \textit{Classical
Dynamics: A Modern Perspective, }Wiley, New York, 1974.

\bibitem{sundermeyer}  K. Sundermeyer, \textit{Constrained Dynamics, }(Lect.
Notes Phys. vol.169) Berlin, Heidelberg, New York: Springer 1982.

\bibitem{qed}  \textit{\ }M.~A.~Santos, J.~C.~de Mello and P.~Pitanga, 
\textit{Physical Variables in Gauge Theories, } \textit{Z.\ Phys.}\ C 
\textbf{55} (1992)271.
%\cite{mmi}
\bibitem{mmi}  M.~A.~De Andrade, M.~A.~Santos and I.~V.~Vancea, \textit{%
Unconstrained Variables of Non-Commutative Open Strings, JHEP} \textbf{0106} 
026 (2001) [hep-th/0104154].
%%CITATION = JHEPA,0106,026;%%
\bibitem{c-lee}  M.A. Santos, J.C. de Mello and P. Pitanga, \textit{The
Christ-Lee Model in the Approach of the Symplectic Projector Method, Braz.
J. Phys.} \textbf{23} (1993) 214.
%\cite{revmmi}
\bibitem{revmmi}  M.~A.~De Andrade, M.~A.~Santos and I.~V.~Vancea, \textit{%
Local Physical Coordinates from Symplectic Projector Method, Mod.Phys.Lett.}
{\textbf A16} (2001) 1907 [\textit{hep-th}/0108197].
%%CITATION = MPLAE,A16,1907;%%
\bibitem{helayel}  M.~A.~Santos and J.~A.~Helayel-Neto, \textit{Physical
Variables for the Chern-Simons-Maxwell Theory without Matter, } \textit{%
hep-th}/9905065.

\bibitem{helayel2} M.~A.~Santos, J.~A.~Helayel-Neto and P. I. Trajtenberg, 
{\it Probing the Symplectic Projector Method of Quantization in Planar Gauge Models},
painel apresentado no XXII Encontro Nacional de F\'{\i}sica de Part\'{\i}culas e Campos,
S\~ao Louren\c co - MG, 2001; submetido para publica\c c\~ao.

\bibitem{alvaro}  L.~R.~Manssur, A.~L.~Nogueira and M.~A.~Santos, 
\textit{An Extended Abelian Chern-Simons Model and the Symplectic Projector
Method, hep-th}/0005214.

\bibitem{christ-lee}  N. H. Christ and T. D. Lee, \textit{Phys. Rev. }%
\textbf{D22 }(1980) 939.

\bibitem{costa-girotti}  M. E .V. Costa and H. O. Girotti, \textit{Phys.
Rev. }\textbf{D24} (1981) 3323.

\bibitem{phokhorov}  L. V. Phokhorov, \textit{Sov. J. Nucl. Phys}. \textbf{%
35 }(1982) 129.

\bibitem{bouzas}  A. Bouzas, L. N. Epele, H. Fanchiotti and C. A. Garcia
Canal, \textit{Phys. Rev. }\textbf{A42 }(1990) 90.

\bibitem{bigatti}  D. Bigatti and L. Susskind, \textit{Phys. Rev.} \textbf{%
D62} (2000)[\textit{hep-th/}9908056].

\bibitem{sakurai}  J. J. Sakurai, \textit{Advanced Quantum Mechanics, }%
Addison-Wesley, 1967.

\bibitem{schwinger}  J. Schwinger, \textit{Phys. Rev. }\textbf{D}128 (1962)
2425.

\bibitem{swieca}  J. L\"{o}wenstein and J. A. Swieca, \textit{Ann. Phys. (}%
N.Y.) 68 (1971) 172.

\bibitem{kogut}  J. B. Kogut and L. S\"{u}sskind, \textit{Phys. Rev. }%
\textbf{D}13 (1975) 2187.

\bibitem{clovis}  C. Wotzasek, \textit{Acta Phys. Pol. }\textbf{B}21 (1990)
463.

\bibitem{devecchi}  F. P. Devecchi, M. Fleck, H. O. Girotti, M. Gomes and A.
J. da Silva, \textit{Ann. Phys. }\textbf{242 }(1995) 275.

\bibitem{kr}  M. Kalb and P. Ramond, \textit{Phys. Rev}. \textbf{D9 }(1974)
2273.

\bibitem{cs}  E. Cremmer and J. Sherk, \textit{Nucl. Phys}. \textbf{B72 }%
(1974) 117.

\bibitem{cchmn1}  H. R. Christiansen, M. S. Cunha, J. A. Hela\"{y}el-Neto, L.
R. U. Mansur and A. L. M. A. Nogueira, \textit{I. J. Mod. Phys}. \textbf{A14}
(1999) 147 (hep-th/9802096).

\bibitem{cchmn2}  H. R. Christiansen, M. S. Cunha, J. A. Hela\"{y}el-Neto, L.
R. U. Mansur and A. L. M. A. Nogueira, \textit{I. J. Mod. Phys.} \textbf{A14 
}(1999) 1721 (hep-th/9805128).

\bibitem{witten}  E. Witten, \textit{Nucl. Phys.} \textbf{B460} (1996) 335,
hep-th/9510135.

\bibitem{seiberg}  N. Seiberg and E. Witten, \textit{JHEP} \textbf{09} 
(1999) 032, hep-th/9908142.

\bibitem{ardalan1}  F. Ardalan, H. Arfaei and M. M. Sheikh-Jabbari, \textit{%
JHEP} \textbf{02}(1999) 016, hep-th/9810072.

\bibitem{ardalan2}  F. Ardalan, H. Arfaei and M. M. Sheikh-Jabbari, \textit{%
Nucl. Phys.} \textbf{B576} (2000) 578, hep-th/9906161.

\bibitem{sheik}  M. M. Sheikh-Jabbari and A. Shirzad, \textit{Eur. Phys. J.} 
\textbf{C19} (2001) 383, hep-th/9907055.

\bibitem{chu1}  C.-S. Chu and P.-M. Ho, \textit{Nucl. Phys}. \textbf{B550} 
(1999) 151, hep-th/9812219.

\bibitem{chu2}  C.-S. Chu and P.-M. Ho, \textit{Nucl. Phys}. \textbf{B568} 
(2000) 447, hep-th/9906192.

\bibitem{lee}  T. Lee, \textit{Phys. Rev}. \textbf{D62} (2000) 024022,
hep-th/9911140.

\bibitem{zab}  M. Zabzine, \textit{JHEP }\textbf{10} (2000) 042,
hep-th/0005142.

\bibitem{kim}  W. T. Kim and J. J. Oh, \textit{Mod. Phys. Lett}. \textbf{A15} 
(2000) 1597, hep-th/9911085.

\bibitem{rudy}  I. Rudychev, \textit{JHEP} \textbf{04} (2001) 015,
hep-th/0101039. 
\end{thebibliography}
\end{document}